\documentclass[oneside,11pt]{Latex/Classes/PhDthesisPSnPDF}

\newcommand{\figuremacro}[3]{
	\begin{figure}[htbp]
		\centering
		\includegraphics[width=1\textwidth]{#1}
		\caption[#2]{\textbf{#2} - #3}
		\label{figure:#1}
	\end{figure}
}

\usepackage[T1]{fontenc}
\usepackage{array}
\usepackage{pdfpages}
\usepackage{amsthm}

\makeatletter
\renewcommand{\@chapapp}{}
\newenvironment{chapquote}[2][2em]
  {\setlength{\@tempdima}{#1}%
   \def\chapquote@author{#2}%
   \parshape 1 \@tempdima \dimexpr\textwidth-2\@tempdima\relax%
   \itshape}
  {\par\normalfont\hfill--\ \chapquote@author\hspace*{\@tempdima}\par\bigskip}
\makeatother

\theoremstyle{definition}
\newtheorem{definition}{Definition}[section]

\theoremstyle{remark}
\newtheorem*{remark}{Remark}

\usepackage[colorinlistoftodos,prependcaption,textsize=tiny]{todonotes}

\usepackage{graphicx}
\begin{document}

\thispagestyle{empty}

\begin{center}

Vrije Universiteit Amsterdam \hspace*{2cm} Universiteit van Amsterdam

\vspace{1mm}

\hspace*{-7.5cm}\includegraphics[height=25mm]{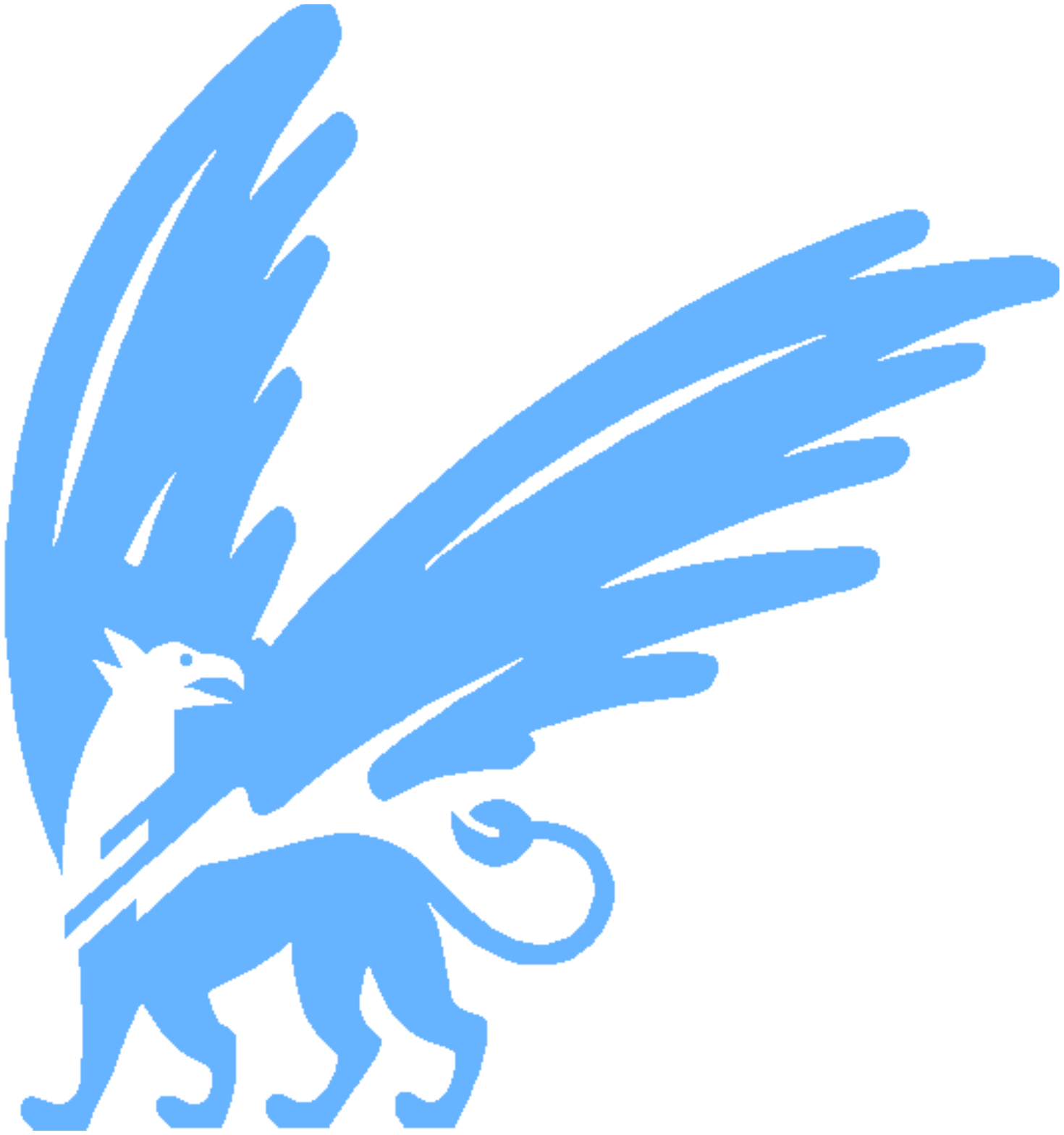}

\vspace*{-2cm}\hspace*{7.5cm}\includegraphics[height=15mm]{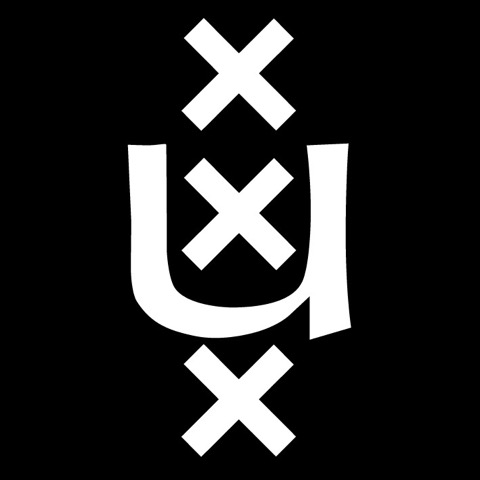}

\vspace{2cm}

{\Large Master's Thesis}

\vspace*{1.5cm}

\rule{.9\linewidth}{.6pt}\\[0.4cm]
{\huge \bfseries SonicChain: A Wait-free, Pseudo-Static Approach Toward Concurrency in Blockchains\par}\vspace{0.4cm}
\rule{.9\linewidth}{.6pt}\\[1.5cm]

\vspace*{1.8mm}

{\Large
\begin{tabular}{l}
{\bf Author:} Kian Paimani (2609260)
\end{tabular}
}

\vspace*{1cm}

\begin{tabular}{ll}
{\it 1st supervisor:}   & dr. ir. A.L. Varbanescu \\
{\it 2nd reader:}       & dr. F. Regazzoni
\end{tabular}

\vspace*{2.5cm}

\textit{A thesis submitted in fulfillment of the requirements for\\ the Master of Science degree in Parallel and Distributed Computer Systems}

\vspace*{1.8cm}

\today\\[4cm] 

\end{center}

\newpage


\hbadness=10000
\hfuzz=50pt


\renewcommand\baselinestretch{1}
\baselineskip=18pt plus1pt


\begin{center}
\textit{``In the future, trusting an opaque institution, a middleman, merchant or intermediary with our interest, would be as archaic a concept, as reckoning on abacuses today'' \\ {\em } -- Dr. Gavin Wood}
\end{center}










\chapter*{Prelude: Web 3.0}

\begin{chapquote}{Web3 Summit 2019 - Berlin}
	My wife asked me why I spoke so softly in the house. I said I was afraid Mark Zuckerberg was
	listening!

	She laughed. I laughed. Alexa laughed. Siri laughed.
\end{chapquote}

I want to briefly recall how I got into the blockchain ecosystem, and how it turned out to be much
more serious and important than what I thought, and why I think you should think the same way. I
will be brief though, there is a lot more to read in this thesis.

I started developing blockchains in 2019, and at first I always admittedly said to my friends and
colleagues that I am interested in it from an engineering perspective. The only other facade that I
knew to blockchains was its monetary aspect. In short, blockchains were two things to me: Complex
distributed systems, and getting rich overnight. I tried to convince myself that I am only
interested in the former.

Soon after, through the works, papers, and talks of visionary pioneers of the blockchain ecosystem,
I learned about the third, perhaps the most important facade of blockchains: That it is merely a
tool in a much larger ecosystem of ideas and beliefs about how we collaborate, perform logistics,
and live as a society. Different people have drawn different ideas here; I will stick to just one
that I find the most relevant: Web 3.0.

Let us look back: web 1.0 was the static web of the '90s and the early years of the 20th century. It
had limited interaction, and only served to show static information. The early web and its
underlying protocols were designed for government and educational purposes. \textit{Trust} is an
absolutely vital asset in web 1.0.

Then came along Mark Zucker... I mean web 2.0. It was dynamic, it could do much more and started
offering way more. We starter building financial and institutional systems on top of it. Except, we
forgot to update any of the protocols. Web 2.0 expanded very quick and very fast, and brought us
here: We can do many great things with web 2.0, it is great. But the data, privacy, and sovereignty
of the user is entirely at the mercy of centralized, giant tech corporations. In a sense, we let web
2.0 expand exponentially, without thinking that its underlying protocols -- the \textit{trust} -- is
still there, and it is becoming more dangerous every day.

It was quite a shock to me to learn that blockchain-enthusiasts are as excited to talk about Satoshi
Nakatomo and the daily bitcoin price, as they are to talk about the Snowden revolution, privacy,
advertising, and democracy. And this is where web 3.0 comes into play. Web 3.0, essentially, is
supposed to eliminate the \textit{trust} embedded in the web 2.0 protocols, allowing users to
interact with one another in a trust-less, decentralized, and more secure manner, using protocols
that are governed transparently, rather than being owned by a big cooperation. And in this picture,
blockchain is really just the tip of the iceberg, just one piece of the puzzle. A much wider army of
tools, protocols and science is needed to develop and build web 3.0. This ranges from bare-bone
sciences such as probability and cryptography, to social matters such as privacy, identity, and
transparent governance of protocols, all the way to the typical stuff: user-friendly interfaces and
APIs for other developers to build upon.

I learned over time that blockchains are not as impressive as they seem. Rather it is the underlying
aspiration that makes it truly exciting. I hope you feel similar someday, and this thesis might help
you learn some of the basics of blockchain along the way.

\begin{abstracts}

Blockchains have a two-sided reputation: they are praised for disrupting some of our institutions
through innovative technology for good, yet notorious for being slow and expensive to use. In this
work, we tackle this issue with concurrency, yet we aim to take a radically different approach by
valuing \textit{simplicity}. We embrace the simplicity through two steps: first, we formulate a
simple runtime mechanism to deal with conflicts called \textit{concurrency delegation}. This method
is much simpler and has less overhead, particularly in scenarios where conflicting transactions are
relatively rare. Moreover, to further reduce the number of conflicting transactions, we propose
using static annotations attached to each transaction, provided by the programmer. These annotations
are pseudo-static: they are static with respect to the lifetime of the transaction, and therefore
are free to use information such as the origin and parameters of the transaction. We propose a
distributor component that can use the output of this pseudo-static annotations and use them to
effectively distribute transactions between threads in the \textit{least}-conflicting way. We name
the outcome of a system combining concurrency delegation and pseudo-static annotations as
\textit{SonicChain}. We evaluate SonicChain for both validation and authoring tasks against a common
workload in blockchains, namely, balance transfers, and observe that it performs expectedly well
while introducing very little overhead and additional complexity to the system.

\end{abstracts}


\frontmatter




\begin{acknowledgements}

I learned all of the engineering and blockchain knowledge needed to conduct this research through
Parity Technologies, a pioneering company developing a wide variety of technologies for the
decentralized future. I joined parity with almost no experience in such fields, and find myself
extremely lucky that I was given the opportunity to learn from their engineers, make mistakes and
grow along the way. My work with Parity will continue after this thesis is over, and I hope for it
to last for many more years.

I took two courses in 2018 at UvA (even though I technically study at VU) about concurrency, both
supervised by Dr. Ana Varbanescu. Both were so enticing that they sparked my interest in concurrency
and performance engineering, eventually giving me the idea to mix it with blockchains for my final
thesis. I recall the first meeting that I had with her about my thesis, and to this day I am
grateful that she accepted and supported my work from that very first day. Looking back at it in
hindsight, \textit{I} probably wouldn't have accepted myself back then if I was in her shoe, give
that I had a very vague idea of what my main idea is, and the rather unorthodox domain of the
thesis. I can only hope to continue learning from Dr. Ana Varbanescu and enjoy it, as I always have.

\end{acknowledgements}


\setcounter{secnumdepth}{3} 
\setcounter{tocdepth}{2}    
\tableofcontents            


\listoffigures	
\listoftables  


%


\markboth{
	\MakeUppercase{\nomname}
}{\MakeUppercase{\nomname}}


\begin{footnotesize}  

\printnomenclature 
\label{nom} 

\end{footnotesize}


\mainmatter

\renewcommand{\chaptername}{} 



\chapter{Introduction} \label{chap:intoroduction}

\begin{chapquote}{Unknown.}
``If Bitcoin was the calculator, Ethereum was the ENIAC\footnote{the first generation computer
developed in 1944. It fills a 20-foot by 40-foot room and has 18,000 vacuum tubes.}. It is
expensive, slow, and hard to work with. The challenge of today is to build the \textbf{commodity,
accessible and performant} computer.''
\end{chapquote}

Blockchains are indeed an interesting topic in recent and coming years. Many believe that they are a
revolutionary technology that will shape our future societies, much like the internet and how it has
impacted many aspects of how we live in the last few decades \cite{pirrongWillBlockchainBe2019}.
Moreover, they are highly sophisticated and inter-disciplinary software artifacts, achieving high
levels of decentralization and security, which was deemed impossible so far. To the contrary, some
people skeptically see them as controversial, or merely a "hyped hoax", and doubt that they will
ever deliver much \textit{real} value to the world. Nonetheless, through the rest of this chapter
and this work overall, we provide ample reasoning to justify why we think otherwise.

In very broad terms, a blockchain is a tamper-proof, append-only ledger that is being maintained in
a decentralized fashion and can only be updated once everyone agrees upon that change as a bundle
of transactions. This bundle of transactions is called a \textbf{block}. Once this block is agreed
upon, it is appended (aka. \textit{chained}) to the ledger, hence the term
\textit{block}-\textit{chain}. Moreover, the ledger itself is public and openly accessible to
anyone. This means that everyone can verify and check the final state of the ledger, and all the
transactions and blocks in its past that lead to this particular ledger state, to verify everything.
At the same time, asymmetric cryptography is the core identity indicator that one needs to interact
with the chain, meaning that one's personal identity can stay private in principle, if that is
desired based on the circumstances. For example, one's public key is not revealing their personal
identity, whilst using names and email addresses does, like in many traditional systems.

In some sense, blockchains are revolutionary because they remove the need for \textit{trust}, and
release it from the control of one entity (e.g. a single bank, an institute, or simply Google), by
encoding it as a self-sovereign decentralized software. Our institutions are built upon the idea
that they manage people's assets, matters and belongings, and they ensure veracity, \textit{because
we trust in them}. In short, they have \textbf{authority}. Of course, this model could work well in
principle, but it suffers from the same problem as some software do: it is a \textbf{single point of
failure}. A corrupt authority is just as destructive as a flawed single point of failure in a
software is. Blockchain attempts to resolve this by deploying software (i.e. itself) in a
transparent and decentralized manner, in which everyone's privacy is respected, whilst at the same
time everyone can still check the veracity of the ledger, and the software itself. In other words,
no single entity should be able to have full control over the system.

Now, all of these great properties do not come cheap. Blockchains are extremely complicated pieces
of software, and they require a great deal of expertise to be written. Moreover, many of the
machinery used to create this \textit{decentralized} and \textit{public} append-only ledger requires
synchronization, serialization, or generally other procedures, that are likely to decrease the
throughput at which the chain can process transactions. This, to some extent, contributes to the
skepticism about blockchains' feasibility. For example, Bitcoin, one of the famously deployed
blockchains to date, consumes a lot of resources to operate, and cannot execute more than around
half a dozen transactions per second \cite{gervaisSecurityPerformanceProof2016}.

Therefore, it is a useful goal to investigate the possibilities through which a blockchain system
can be enhanced to operate \textit{faster}, effectively delivering a higher throughput of
transactions per some unit of time.

\section{Research Questions} \label{chap_intro:sec:resarch_q}

We have seen that blockchains are promising in their technology, and unique traits that they can
deliver. Yet, they are notoriously slow. Therefore, we pursue the goal of improving the
\textit{throughput} of a blockchain system. By throughput, we mean the number of successful
transactions that can be processed per some unit of time. There are numerous ways to
achieve this goal, ranging from redesigning internal protocols within the blockchain to applying
concurrency. In this thesis, we precisely focus on the latter, enabling concurrency within
transactions that are processed and then appended to the ledger. Moreover, we do so by leveraging,
and mixing the best attributes of two different realms of concurrency, namely static analysis and
runtime conflict detection \footnote{By static we mean generally anything which is known at the
\textit{compile} phase, and by runtime the \textit{execution} phase}. This approach is better
compared with other alternatives in section \ref{chap_approach:sec:ways_to_speedup}.

Based on this, we formulate the following as our research questions:

 \begin{enumerate}
    \item [\textbf{RQ1}] What approaches exist to achieve concurrent execution of transactions
    within a blockchain system, and improve the throughput?

    \item [\textbf{RQ2}] How could both static analysis and runtime approaches be combined together
    to achieve a new approach with minimum overhead and measurable benefits?

    \item [\textbf{RQ3}] How would such an approach be evaluated against and compared to others?
 \end{enumerate}

\section{Thesis Outline}
The rest of this thesis is organized as follows: In chapter \ref{chap:background} we provide a
comprehensive background on both blockchains and concurrency, the two pillars of knowledge that we
will build upon. Chapter \ref{chap:approach} starts by defining the \textit{requirements} of a
hypothetical system of interest. Then, some of the contemporary literature and the ways that they
fulfill the requirements are mentioned, effectively acting as our mini "related work" section.
Finally, we introduce our own approach and place it next to other approaches for comparison. Chapter
\ref{chap:impl} acts as a mini detour into the implementation of our system design. In chapter
\ref{chap:bench_analysis} we evaluate our design and implementation, finally bringing us to a
conclusion and future work in chapter \ref{chap:conclusion}.


\chapter{Background} \label{chap:background}

\begin{chapquote}{David Chum et. al. - 1990}
``The use of credit cards today is an act of faith on the part of all concerned. Each party is
vulnerable to fraud by the others, and the cardholder, in particular, has no protection against
surveillance.''
\end{chapquote}

In this chapter, we dive into the background knowledge needed for the rest of this work. Two primary
pillars of knowledge need to be covered: blockchains and distributed systems in section
\ref{chap_bg:sec:blockchains} and concurrency, upon which our solution will be articulated, in
section \ref{chap_bg:sec:concurrency}.

\section{Blockchains And Distributed Ledger Technology} \label{chap_bg:sec:blockchains}

In this section, we provide an overview of the basics of distributed systems, blockchains, and their
underlying technologies. By the end of this section, it is expected that an average reader will know
enough about blockchain systems to be able to follow the rest of our work, and understand the
approach proposed in chapter \ref{chap:approach} and onwards.

\subsection{Centralized, Decentralized and Distributed Systems} \label{chap_bg_:subsec:network}

An introduction to blockchain is always entangled with \textit{distributed} and
\textit{decentralized} systems.

A distributed system is a system in which a group of nodes (each having their own processor and
memory) cooperate and coordinate for a common outcome. From the perspective of an outside user, most
often the distributed nature of the system is transparent, and all the nodes can be seen and
interacted with, as if they were \textit{one cohesive system}
\cite{bashirMASTERINGBLOCKCHAINDistributed2018}.

Indeed, some details differ between the distributed systems and blockchains, yet the underlying
concepts resonate in many ways \cite{herlihyBlockchainsDistributedComputing2019}, and blockchains
can be seen as a form of distributed systems. Like a distributed system, a blockchain also consists
of many nodes, operated either by organizations, or by normal people with their commodity computers.
Similarly, this \textit{distribution} trait is transparent to the end-user when they want to
interact with the blockchain, and they indeed see the system as one cohesive unit. All of these
nodes together form the \textit{blockchain network}.

\nomenclature{Node}{A single entity in a network}
\nomenclature{Blockchain Network}{The set of entire nodes holding (a partial) copy of a particular ledger}

Blockchains are also \textbf{decentralized}. This term was first introduced in a revolutionary paper
in 1964 as a \textbf{middle ground} between purely \textit{centralized} systems that have a single
point of failure, and 100\% \textit{distributed} systems, which are like a mesh (all nodes having
links to many other nodes \cite{baranDistributedCommunicationsNetworks1964} \footnote{The design of
Paul Baran, author of \cite{baranDistributedCommunicationsNetworks1964}, was first proposed, like
many other internet-related technologies, in a military context. His paper was a solution to the
USA's concern about communication links in the aftermath of a nuclear attack in the midst of the
cold war \cite{monicaPaulBaranOrigins}.}). A decentralized system falls somewhere in between, where
no single node's failure can irrecoverably damage the system, and communication is somewhat
distributed, where some nodes might act as hops between different sub-networks.

Blockchains, depending on the implementation, can resonate more with either of the above terms. Most
often, from a networking perspective, they are much closer to the ideals of a distributed system.
From an operational and economical perspective, they can be seen more as decentralized, where the
operational power (i.e. the \textit{authority}) falls into the hands of no single entity, yet a
large enough group of authorities.

\figuremacro
	{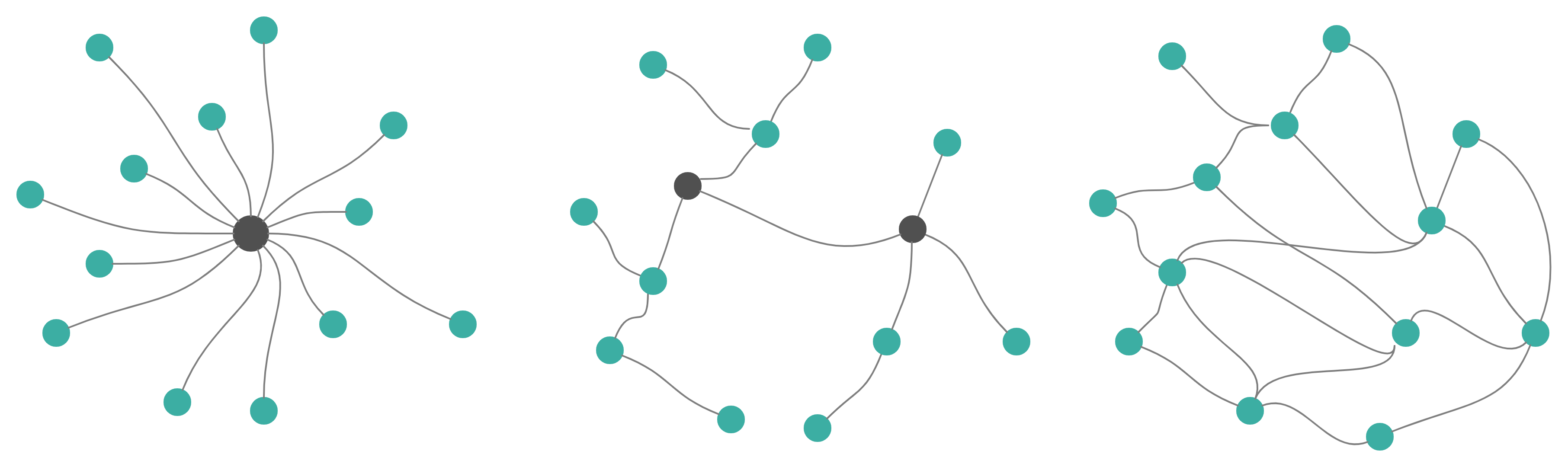} {Types of Networks.} {From left to right: Centralized, Decentralized, and
	Distributed.}

\subsection{From Ideas to Bitcoin: History of Blockchain} \label{chap_bg_:subsec:hisotry}

While most people associate the rise of blockchains with Bitcoin, it is indeed incorrect, because
the basic ideas of blockchains were mentioned decades earlier. The first relevant research paper was
already mentioned in \ref{chap_bg_:subsec:network}. Namely, in
\cite{baranDistributedCommunicationsNetworks1964}, besides the definition of a decentralized system,
the paper also describes many metrics regarding how secure a network should be, under certain
attacks.

Next, \cite{diffieNewDirectionsCryptography1976} famously introduced what is know as Diffie-Hellman
Key Exchange, which is the backbone of public-key encryption. Moreover, this key exchange is heavily
inspired by \cite{merkleSecureCommunicationsInsecure1978}, which depicts more ways in which
cryptography can be used to secure online communication. Together, these papers form the
\textit{digital signature scheme}, which is heavily used in all blockchain systems \footnote{Many of
these works were deemed military applications at the time, hence the release dates are what is
referred to as the "public dates", not the original, potentially concealed dates of their
discovery.}.

Moreover, the idea of blockchain itself predates Bitcoin. The idea of chaining data together, whilst
placing some digest of the previous piece (i.e. a \textit{hash} thereof) in the header of the next
one was first introduced in \cite{haberHowTimestampDigital1991}. This, in fact, is exactly the
underlying reason that a blockchain, as a data structure, can be seen as an append-only,
tamper-proof ledger. Any change to previous blocks will break the hash chain and cause the hash of
the latest block to become different, making any changes to the history of the data structure
identifiable, hence \textit{tamper-proof}.

Finally, \cite{chaumUntraceableElectronicCash1990} introduced the idea of using digital computers as
a means of currency in 1990, as an alternative to the rise of credit cards at the time. There were a
number of problems with this approach, including the famous double-spend problem, in which an entity
can spend one unit of currency numerous times. Finally, in 2008, an unknown scientist who used the
name Satoshi Nakatomo released the first draft of the Bitcoin whitepaper. In his work, he proposed
Proof of Work as a means of solving the double-spend problem, among other details and improvements
\cite{nakamotoBitcoinPeertoPeerElectronic}. Note that the idea of Proof of Work itself goes back,
yet again, to 1993. This concept was first introduced in
\cite{dworkPricingProcessingCombatting1993}, as means of spam protection in early email services.

\subsection{1000 Feet View of a Blockchain} \label{chap_bg:subsec:100ft}

\figuremacro
	{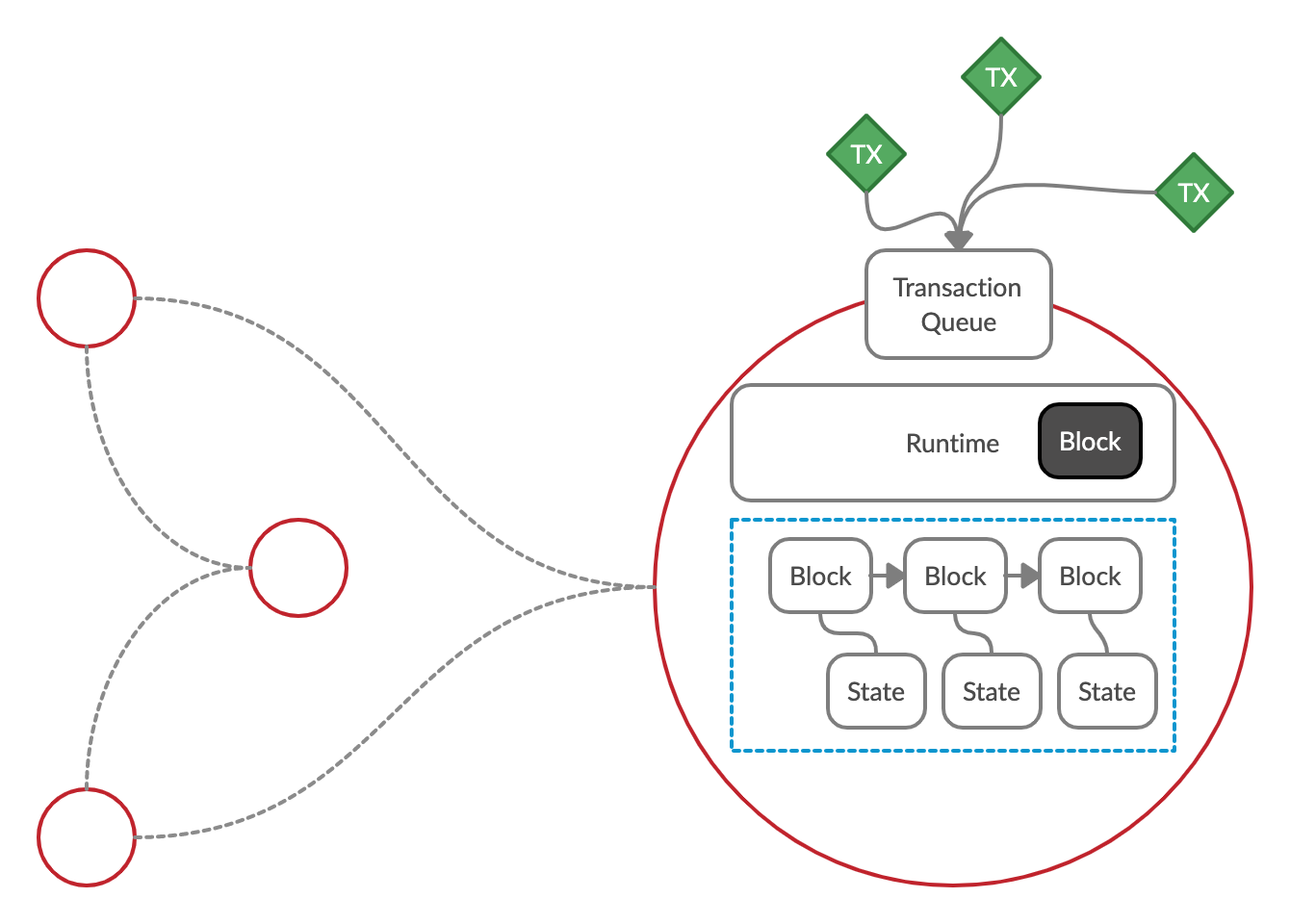} {A 100ft View of the Blockchain Network} {Nodes (red circles), Ledger state
	(dashed green box), Transaction (green diamonds).}

Before going any further, we provide this, so-called, 1000 feet view of the blockchain. This small
section provides the big picture in the blockchain world, while the upcoming sections go into the
fine depth thereof. This section introduces a lot of jargon all at once, but will help comprehend
the rest of the chapter.

Figure \ref{figure:figures/100ft.png} shows a very broad overview of a blockchain network and the
processes within it. Each red circle is a node in the decentralized network. We zoom into only one
of them. Each node holds two very important components: A runtime, and a ledger (dashed blue box).

The \textbf{ledger} is composed of the chain of blocks, and some auxiliary \textbf{state} for each
block. These two are basically the entire view of the world. The blocks are linked together by a
hash chain. We see a block outside of the state (black block). This is a candidate block, one that
this node might propose to the rest of the network at some point to be appended.

The \textbf{runtime} is simply the core logic of the chain. This core logic, the runtime, has the
responsibility of \textit{updating} the ledger based on \textit{transactions}. Transactions (green
diamonds) are sent to a node from the outer world, potentially from end-users. The transactions are
kept in a separate data structure (transaction queue) until getting a chance at being put into a
block.

\subsection{Preliminary Concepts} \label{chap_bg:sec:preliminary}

Having known where the blockchain's idea originates from, and which fields of previous knowledge in
the last half a century it aggregates, we can now have a closer look at these technologies and
eventually, build up a clear and concrete understanding of what a blockchain is and how it works.

\subsubsection{Elliptic Curve Cryptography} \label{chap_bg:subsec:ecc}

We mentioned the Diffie-Hellman key exchange scheme in section \ref{chap_bg_:subsec:hisotry}. Key
exchange is basically a mechanism to establish a \textit{symmetric} key, using only
\textit{asymmetric} data. In other words, two participants can come up with a common shared
symmetric key (used for encrypting data) without ever sharing it over the network\footnote{Readers
may refer to \cite{diffieNewDirectionsCryptography1976} for more information about the details of
how this mechanism works.}. Indeed, while the underlying principles are the same, for better
performance, most modern distributed systems work with another mechanism that is the more advanced
variant of Diffie-Hellman, namely Elliptic Curve Cryptography (ECC). Elliptic Curves offer the same
properties as Diffie-Hellman, with similar security measures, whilst being faster to compute and
needing smaller key sizes. A key exchange, in short, allows for \textbf{asymmetric cryptography},
the variant of cryptography that needs no secrete medium to exchange initial keys, and therefore is
truly applicable to distributed systems. In asymmetric cryptography, a key \textit{pair} is
generated at each entity. A \textbf{public} key, which can be, as the name suggests, publicly shared
with anyone, and a \textbf{private} key that must be kept secret. Any data signed with the private
key can be verified using the public key. This allows for integrity checks, and allows anyone to
verify the \textit{origin} of a message. Hence, the private key is also referred to as the
\textbf{signature}. Moreover, any data encrypted with the public key can only be decrypted with the
private key. This allows confidentiality.

Many useful properties can be achieved using asymmetric cryptography, and many massively useful
applications adopt it \footnote{The device that you are using to read this line of text has probably
already done at least one operation related to asymmetric cryptography since you started reading
this footnote. This is how relevant they \textit{really} are.}. For blockchains, we are particularly
interested in \textbf{signatures}. Signatures allow entities to verify the integrity and the origin
of any message. Moreover, the public portion of a key, i.e. a public key, can be used as an
identifier for an entity.

For example, in the context of banking, a public key can be seen as an account number. It is public
and known to everyone, and knowing it does not grant anyone the authority to withdraw money from an
account. The private key is the piece that gives one entity \textit{authority} over an account, much
like your physical presence at the bank and signing a paper, in a traditional banking system. This
is a very common pattern in almost all blockchain and distributed systems: using private keys to
sign messages, and using public keys as identities.

RSA and DSA are both non-elliptic signature schemes that are commonly known to date. ECDSA, short
for \textbf{E}lliptic \textbf{C}ureve DSA, is the Elliptic Curve variant of the latter. Albeit,
ECDSA is a subject of debate, due to its proven insecurities \cite{brumleyRemoteTimingAttacks2011},
and its performance. Hence, more recent, non-patented and open standard \footnote{Unlike ECDSA which
is developed and patented by NIST, which in fact is the reason why many people doubt its security.}
curves, such as EdDSA, are the most commonly used. EdDSA, short for Edwards-curve Digital Signature
Algorithm is based on the open standard Edward-curve and its reference, parameters, and
implementation are all public domain.

All in all, cryptography, and specifically digital signatures, play an integral role in the
blockchain technology, and allow it to operate in a distributed way, where external messages can be
verified from their signatures.

\subsubsection{Hash Functions} \label{chap_bg:subsec:hash}

Hash functions, similar to elliptic curve cryptography, are among the mathematical backbones of
blockchains. A hash function is basically a function that takes some bits of data as input and
returns some bits of output in return. All hash functions have an important property: they produce a
\textbf{fixed sized output}, regardless of the input size. Also, a hash function ensures that
changing anything in the input, as small as one bit, does result in an entirely different output.

Given these properties, one can assume that the hash of some piece of data can be seen as its
\textbf{digest}. If the hash of two arbitrarily large pieces of data is the same, you can assume
that their underlying data are indeed the same. This is quite helpful to ensure that some cloned
data is not tampered with, in a distributed environment. If we only distribute the hash of the
original copy in a secure way, everyone can verify that they have a correct clone, without the need
to check anything else.

Albeit, a secure hash function needs to provide more properties. First, the hash function needs to
ensure that no two different inputs can lead to the same hash. This - the situation that different
inputs generate the same hash - is called a \textit{collision}, and the probability of collision in
a hash function should be sufficiently low for it to be secure. Moreover, a hash function must be a
\textit{one way} function, meaning that it cannot be reversed in a feasible manner. By feasible we
effectively mean \textbf{timely}: if reversing a function takes a few million years, it is
\textit{possible}, but not \textit{feasible}. The entire security of hash functions and digital
signatures is based on the fact that breaking them is not feasible \footnote{And, yes, we are aware
of quantum computing, but that is a story for another day.}. So, given some hash output, one cannot
know the input that leads to that hash. Hash functions that have this property are typically called
\textit{cryptographic} hash functions. Cryptographic hash functions are commonly used, next to
asymmetric cryptography, for authentication and integrity checks, where the sender can sign only a
hash of a message and send it over the network, such as in the common \textbf{M}essage
\textbf{A}uthentication \textbf{C}ode, pattern \cite{bellareKeyingHashFunctions1996} (MAC).

\subsubsection{Peer to Peer Network} \label{chap_bg:subsec:p2p}

From a networking perspective, a blockchain is a purely peer to peer distributed network. A peer to
peer network is one in which many nodes form a mesh of connections between them, and they are more
or less of the same role and privilege.

A peer to peer network is the architectural equivalent of what was explained as a
\textbf{distributed} network earlier in this chapter. The opposing architecture is what is known as
the \textit{client-server} model, in which one node is the server and everyone else is a client. In
other words, the client-server model is \textbf{centralized}, in the sense that only the server
contains valuable resources (whatever that resource might be: computation power, data, etc.) and
serves it to all other clients.

Unlike a client-server model, a peer to peer network does not have a single point of failure: there
is no notion of client and server, and all of the entities have the same role, being simply called
\textit{nodes}. Having no servers to serve some data, it is straightforward to say that peer to peer
networks are \textit{collaborative}. A node can consume some resources from another node by
requesting data from it, whilst being the producer for another node by serving data to it. This is
radically different from the traditional client-server model, in which the server is always the
producer and clients are only consumers, and effectively have no control over the data that they are
being served.

Each node in a peer to peer network is constantly talking to other neighboring nodes. For this, they
establish communication links between one another. Regardless of the transport protocol (TCP, QUIC
\cite{carlucciHTTPUDPExperimental2015}, etc.), these connections must be secure and encrypted for
the network to be resilient. Both elliptic curve cryptography and hash functions explained in the
previous sections, provide the technology needed to achieve this.

In the rest of this work, we are particularly interested in the fact that in a blockchain system,
the networking layer provides \textit{gossip} capability. The gossip protocol is an epidemic
procedure to disseminate data to all neighboring nodes, to eventually reach the entire network. In a
nutshell, it is an \textit{eventually consistent} protocol to ensure that some messages are being
constantly gossiped around, until eventually everyone sees them. Blockchains use the gossip protocol
to propagate the messages that they receive from the end-user (the most important of which being
transactions, explained in \ref{chap_bg:subsec:transaction_sig}).

A distributed system must be seen as a cohesive system from outside. Thus, a transaction that a user
submits to one node of the network should have the same chance of being appended to the ledger by
any of the nodes in the future. Therefore, it must be gossiped around. This becomes more clear when
we discuss block authoring in section \ref{chap_bg:subsec:consensus_authorship}.

\subsubsection{Key-Value Database} \label{chap_bg:subsec:kvdb}

Shifting perspective, a blockchain is akin to a distributed database with very high redundancy,
namely one copy per node. One might argue that this is too simplistic, but even the brief
description that we have already provided (see \ref{chap_bg:subsec:100ft}) commensurates with this.
Transactions can be submitted to a blockchain. These transactions are then added to a bundle, called
a block, which is then chained with all the previous blocks, forming a chain of blocks. All nodes
maintain their view of this chain of blocks, and basically, that is what the blockchain is: a
database for storing some chain of blocks.

Now we can explain why in figure \ref{figure:figures/100ft.png} each block was linked with some
auxiliary data. Next to the block database, most blockchains store other data as well, to facilitate
more complex logic. For example, in Bitcoin, that logic needs to maintain a list of accounts and
balances, and perform basic math on top of them\footnote{In reality, Bitcoin does something slightly
different, which is known as the UTXO model, which we omit to explain here for simplicity
\cite{delgado-seguraAnalysisBitcoinUTXO2017}.}. To know an account's balance, it is infeasible to
re-calculate it every time from the known history of previous transactions. That would be equivalent
to an ATM machine re-executing all your previous transactions to know your current balance every
time you use it. Thus, we need some sort of database as well, to store the auxiliary data that the
blockchain logic needs - like, the list of accounts and their balances in our example. This
auxiliary data is called the \textbf{state}, and is usually implemented in the form of a key-value
database.

A key-value database is a database that can be queried similar to a \textit{map}. Any value inserted
in the database needs to be linked with a \textit{key}. This value is then placed in conjunction
with that key. The same key can be used to retrieve, update, or delete the value. For example, in a
bitcoin-like system, the keys are account identifiers (which we already mentioned are most often
just public cryptographic keys), and the values are simply the account balances, some numeric value.

Indeed, a more complicated blockchain, that does more than simple accounting, will have a more
complicated state layout. Even more, chains that support the execution of arbitrary code (e.g. Smart
Contracts \cite{EthereumWhitepaper}), like Ethereum, allow any key-value data pair to be inserted
into the state.

One challenge for nodes in a blockchain network is to keep a \textbf{persistent} view of the state.
For example, Alice's view of how much money Bob owns needs to be the same as everyone else's view.
But, before we dive into this aspect, let us first formalize the means of \textit{updating the
state}: the \textbf{transactions}.

\subsubsection{Transactions and Signatures} \label{chap_bg:subsec:transaction_sig}

\nomenclature{Transaction}{Arbitrary data provided by the outer world that can update the blockchain's state}
\nomenclature{Origin}{The source (i.e. the sender) of a transaction. Usually Provided through public-key cryptography}

Transactions are pieces of information submitted to the system, that are eventually appended to the
blockchain in the form of a new block. And, as mentioned, everyone keeps the history of all blocks,
essentially having the ability to replay the history and make sure that an account claiming to have
a certain number of tokens\footnote{Tokens are the equivalent of a monetary unit of currency, like a
coin in the jargon of digital money.} does indeed own it.

The concept of \textit{state} is the main reason why transactions exist. Transactions most often
cause some sort of \textit{update} to happen in the state. Moreover, transactions are accountable,
meaning that they most often contain a signature of their entire payload, to ensure both integrity
and accountability. For example, if Alice wants to issue a \texttt{transfer} transaction to send
some tokens to Bob, the chain will only accept this transaction if is signed with by Alice's private
key. Consequently, if there is a fee associated with this transfer, it is deducted from Alice's
account. This is where the link between identifiers and public keys also becomes more important.
Each transaction has an \textit{origin}, which is basically the identifier of the entity which sent
that transaction. Each transaction also has a signature, which is basically the entire (or only the
sensitive parts of the) payload of the transaction, signed with the private key associated with the
aforementioned origin. Indeed, a transaction is valid only if the signature and the public key
(i.e., the \textit{origin}) match.

This is a radically new usage of public-key cryptography in blockchains, where one can generate a
private key using the computational power of a personal machine, and store some tokens linked to it
on a network operated by many decentralized nodes; that private key is the one and only key that can
unlock those tokens and spend them. Although in this chapter we mostly use examples of a
cryptocurrency (e.g. Bitcoin), we should note that an online currency and token transfer is among
the simplest forms of blockchain transactions. Depending on the functionality of a particular
blockchain, its transactions can have specific logic and complexity. Nonetheless, containing a
\textit{signature} and some notion of \textit{origin} is very common for most use cases.

Let us recap some facts from the previous sections:

\begin{itemize}
	\item A blockchain is a peer to peer network in which a transaction received by one node will
	eventually reach other nodes.
	\item Nodes apply transaction to update some \textit{state}.
	\item Nodes need to keep a persistent view of the state.
\end{itemize}

A system with the combination of the above characteristics can easily come to a race conditions. One
ramification of this race condition is the double-spend problem. Imagine Eve owns 50 tokens. She
sends one transaction to Alice, spending 40 tokens. Alice checks that Eve has enough tokens to make
this spend, updates Eve's account balance to 10, basically updating her own view of the state. Now,
if Eve sends the exact same transaction at the same time to Bob, it will also succeed, if Alice and
Bob have not yet had time to gossip their local transactions to one another.

To solve this, blockchains agree on a contract: the state can \textbf{only} be updated via appending
a new block to the known chain of blocks, not one single transaction at a time. This allows to
compensate for potential gossip delays to some extent, and is explained in more detail in the next
section.

\subsubsection{Blocks} \label{chap_bg:subsec:block}

\nomenclature{Block}{A bundle of transaction that can be appended to the chain}
\nomenclature{Parent Hash}{A hash of the previous block, mentioned in the header of all blocks}
\nomenclature{Genesis Block}{The first block of each chain, and it has no parent hash}

Blocks are nothing but bundles of transactions, and they allow for ordering, which somewhat relaxes
the problem explained in the previous section, namely the race condition between transactions. To do
so, blocks allow nodes to agree on some particular \textit{order} to apply transactions.
Specifically, a node, instead of trying to apply transactions that it has received via the gossip
protocol, in \textit{some random order}, will wait to receive a block from other nodes, and then
apply the transactions therein in the same order as stated in the block. This is called block
\textit{import}.

Transactions inside a block are ordered, and applying them sequentially is fully deterministic: it
will always lead to the same result. Moreover, in the example of the previous section, it is no
longer possible for Eve to spend some tokens twice, because a block will eventually force some
ordering of her transactions, meaning that whichever appears second will indeed fail, because the
effects of the first one are already apparent and persistent in the state of any node that is
executing the block. In other words, as long as a node imports blocks, the transactions within it
are ordered and the aforementioned race condition cannot happen.

A block also contains a small, yet very important piece of data called \textit{parent hash}. This is
basically a hash of the entire content of the last know block of the chain. There is exactly one
block in each chain that has no parent, the first block. This is a special case, and is called the
\textit{genesis block}. This, combined with the properties of the hash function explained in
\ref{chap_bg:subsec:hash}, brings about the tamper-proof-ness of all the blocks. In other words, the
history of operations cannot be mutated. For example, if everyone in the network already knows that
the parent hash of the last known block is $H_1$, it is impossible for Eve to inject, remove, or
change any of the previous transaction, because this will inevitably cause the final hash to be some
other value, $H_2$, which in principle should be very different from $H_1$\footnote{In principle,
the probability of collision (the hash of some \textbf{tampered} chain of blocks being the same as
the valid one) is not absolute zero, but it is so small that it is commonly referred to
\textit{astronomically small}, meaning that it will probably take millions of years for a collision
to happen. As a malicious user, you most often don't want to wait that long.}.

All in all, blocks make the blockchain more tamper-proof (at least the history of transactions), and
bring some ordering constraints regarding the order in which transactions need to be applied.
Nonetheless, with a bit of contemplation, one soon realizes that this is not really solving the race
condition, but rather just changing its \textit{granularity}. Instead of the question of which
transaction to apply next, we now have the problem of which block to append next. This is because,
intentionally, we haven't yet mentioned \textit{who} can propose new blocks to be appended, and
\textit{when}. We have only assumed that we \textit{somehow} receive blocks over the network. This
brings us to the consensus and authorship of blocks, explained in the next section.

\subsubsection{Consensus and Block Authoring} \label{chap_bg:subsec:consensus_authorship}

\nomenclature{Block Authoring}{The sensitive task of preparing a new block and propagating it to the network}

The consensus protocol in a blockchain consists of a set of algorithms that ensure all nodes in the
network maintain an eventually consistent view of the ledger state (both the chain itself, and the
state). The protocol needs to address problems such as emerging network partition, software
failures, and the Byzantine General Problem \cite{lamportByzantineGeneralsProblem1982}, the sate in
which a portion of the nodes in the network \textit{intentionally} misbehave.

For brevity, we only focus on one aspect of the consensus which is more relevant to our work,
namely, the decision of \textit{block authoring}: deciding who can author blocks, and when.

Recall that nodes in a blockchain network form a distributed system. Therefore, each node could have
a different view of the blockchain, and each node might also have a different set of transactions to
build a new block out of (due to the fact that the underlying gossip might have delivered different
transactions to different nodes at a certain point in time). In principle, any of these nodes can
bundle some transactions in a block and propagate it over the network, \textit{claiming} that it
should be appended to the blockchain. This will indeed lead to chaos. To avoid this, block authoring
is a mechanism to dictate who can author the next block\footnote{In some sense, if blockchains are a
democratic system, block authoring is a protocol to chose a \textit{temporary} dictator.}. This
decision must be solved in a decentralized and provable manner. For example, in Proof of Work, each
block must be hashed together with a variable such that the final of the block hash has a certain
number of leading zeros. This is hard to compute, hence the system is resilient against an invalid
block, namely those that were authored without respecting the authoring rules of the consensus
protocol. Moreover, this is provable: any node that receives a candidate block can hash it again and
ensure that the block is valid with respect to Proof of Work. In this thesis, the terms "block
author" and "validators" are used to refer to the entity that proposes the candidate block, and all
the other nodes that validate it, respectively.

\begin{definition} \label{def:auhtor_validator}

	\textit{Authoring and Validating}.

	\textbf{Author}: the network entity that proposes a new candidate block. This task is called
	block \textit{authoring}.

	\textbf{Validators}: All other nodes who receive this block and ensure its veracity. The act of
	ensuring veracity is called \textit{validating} or \textit{importing} a block.
\end{definition}

In a Proof of Work scheme, the next author is basically whoever manages to solve the Proof of Work
puzzle faster.

\begin{definition} \label{def:pow} Given the adjustable parameter $d$ and a candidate block data
$b$, solving the Proof of Work puzzle is the process of finding a number $n$ such that:

	\begin{equation}
		Hash(b || n) <= d
	\end{equation}

Here, $d$ is usually some power of 2 which is equal to a certain number of leading zeros in the
output.
\end{definition}

Indeed, this is very slow and inefficient, to the point that many have raised concerns even about
the climate impact of the Bitcoin network\footnote{Some estimates show the annual carbon emission of
the Bitcoin network is more than that of Switzerland\cite{stollCarbonFootprintBitcoin2019}.}. There
are other consensus schemes, such as Proof of Stake, combined with verifiable random functions
\cite{dodisVerifiableRandomFunction2005}, that solve the same problem, without wasting a lot of
electricity.

Nonetheless, we can see how this solves the synchronization issue in blockchains. A block serializes
a bundle of transactions. The consensus protocol, namely its block authoring protocol, regulates the
block production, so that not everyone can propose candidate blocks at the same time.

\subsubsection{Interlude: The types of blockchains} \label{chap_bg:subsec:blockchain_types}

So far, we have only talked about \textit{permissionless} blockchains, which are the focus of this
work. Nonetheless, now is a good time to mention that a permissionless blockchain is only one out of
three categories of blockchains:

\begin{itemize}
	\item \textbf{Permissionless} blockchains: A type of blockchain in which no single entity has
	any power over the network. Such networks are called permissionless because one needs no
	permission from any authority to perform an action. For example, as long as one pays the
	corresponding fee, one can always submit a transaction to a permissionless network, i.e. one
	cannot be banned by some authority. Or, one can always decide to be a candidate for block
	authoring, if one wishes to do so. One might not succeed in doing so (e.g. in a Proof of Work,
	weak hardware always fails to find a proper hash in time and essentially waste power with no
	gain), but one has the freedom to do all of these actions. Such blockchains truly adhere to the
	decentralized goals of the blockchain ecosystem.
	\item \textbf{Consortium} blockchains: In this type of blockchains, users can still interact
	with the chain freely, but most \textit{consensus critical} actions are not permissionless. For
	example, a chain might decide to delegate the task of block authoring to a fixed number of
	trusted nodes. In such a scenario, none of the mentioned Proof of Work schemes is needed, and
	the authoring rules can be simplified to a round-robin block authoring. Albeit, such chains are
	questionable because they don't really solve the main problem of making systems trustless. Such
	chains are called Proof of Authority, meaning that a node can author a block by the virtue of
	being a member of a fixed set of authorities \cite{deangelisPBFTVsProofofauthority2018}. And
	from the perspective of the end-user, one must still \textit{trust} in the honesty and goodwill
	of these authorities.
	\item \textbf{Private} blockchains: these blockchains use the same technology as Permissionless
	blockchains to establish trust between organizations, but they are not open to the public. A
	common example would be a chain that maintains government records between different ministries.
\end{itemize}

It is important to note that many aspects of the consensus protocol, including its complexity,
change based on the above taxonomy. The permissionless chains will typically have the most difficult
type of consensus, because ensuring veracity is quite hard in a decentralized environment where
anyone might misbehave. Albeit, the rationale of the decentralization advocates is that by making
the system transparent and open to the public, we actually gain more security comparing to hiding it
behind servers and firewalls\footnote{One reasonably might see this concept resonating with the Open
Source Software movement, where open-source software is claimed to be more secure than a closed
source one.}, because we can also attract more honest participants, who can check the system and
make sure it behaves correctly.

\begin{table}[ht]
\centering
	\caption{Types of blockchain based on consensus.}
	\label{table:blockchain_types}
	\begin{tabular}{lllll}
													& \multicolumn{3}{c}{\textbf{Blockchain Type}} &
													\\
													& Public & Consortium          & Private       &
	\\ \cline{1-4} \multicolumn{1}{l|}{\textbf{Permissionless?}}   & Yes    & No                  &
	No            &  \\
	\multicolumn{1}{l|}{\textbf{Read?}}             & Anyone & Depends             & Invite Only   &
	\\
	\multicolumn{1}{l|}{\textbf{Write?}}            & Anyone & Trusted Authorities & Invite Only   &
	\\
	\multicolumn{1}{l|}{\textbf{Owner}}             & Nobody & Multiple Entities   & Single Entity &
	\\
	\multicolumn{1}{l|}{\textbf{Transaction Speed}} & Slow   & Fast                & Fast          &
	\\ \hline
	\end{tabular}
\end{table}

Due to all this complexity, consensus remains a cutting-edge field of research in the blockchain
ecosystem. In table \ref{table:blockchain_types}, we show how the consensus is also a major factor
in the throughput of the blockchain, which is our metric of interest in this work. This correlation
is later explained in \ref{chap_approach:sec:ways_to_speedup}.

\subsubsection{Forks: A Glitch in The Consensus Protocol}

\nomenclature{Canonical Chain}{The current based chain that most of the nodes agree upon (particularly in the context of a fork)}

Coming back to the \textit{permissionless} block authoring schemes mentioned in
\ref{chap_bg:subsec:consensus_authorship}, it turns out that a perfect consensus cannot exist in a
permissionless network \cite{wangSurveyConsensusMechanisms2019}. Aside from problems such as a node
being malicious and network partitions, there could be other non-malicious scenarios in which
everything in the network is seemingly fine, yet nodes end up with different blockchain views. A
simple scenario that can lead to this is if, by chance, two nodes manage to solve the Proof of Work
puzzle almost at the same time. They both create a \textit{completely valid} block candidate and
propagate it to the network. Some nodes might see one of the candidates first, while the others
might see another one first. Such a scenario is called a \textbf{Fork}: a state in which nodes have
been partitioned into smaller groups, each having their own blockchain views. Most consensus
protocols solve this by adopting a \textit{longest chain} rule. Eventually, once all block
candidates have been propagated, each node chooses the longest chain that they can build, and that
is the accepted one. This chain is called the \textit{canonical chain}, and the last block in it is
called the \textit{best-block} or the \textit{head} of the blockchain. Based on the canonical chain,
the state can also be re-created and stored.

Aside from malicious forks (that we do not cover here), and forks due to decentralization such as
the example above, there could be \textit{federated} forks as well. For example, if a group of nodes
in a blockchain network decide to make a particular change in the history, and they all agree on it,
they can simply fork from the original chain and make their new chain. This new chain has some
common prefix with the original one, but it diverges at some point. A very famous example of this is
the Ethereum Classic fork from Ethereum network \cite{vignaGreatDigitalCurrencyDebate2016}. After a
hack due to a software bug, a lot of funds got frozen in the main Ethereum network. A group of
network participants decided to revert the hack. This was not widely accepted in the Ethereum
ecosystem\footnote{After all, it defies all the \textit{immutability} properties of a blockchain.}
and thus, a fork happened, giving birth to the \textit{Ethereum Classic} network.

\figuremacro{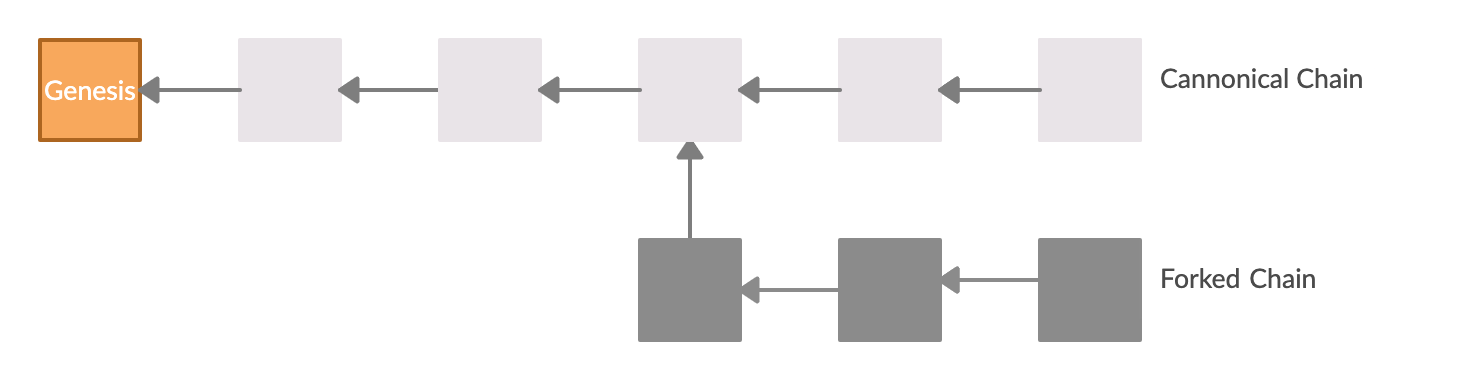}{Forks}{The cannon chain and the forked chain both have a common
prefix, yet have \textit{different} best-blocks.}

\subsubsection{Merkle Tree and State Root} \label{chap_bg:subsec:trie}

Recall from \ref{chap_bg:subsec:kvdb} that blockchains store some sort of \textbf{state} next to
their history of blocks as well. We further explain the reasons for this design in this section. To
recap, the state is a key-value database that represents the state of the world, i.e., all the data
that is stored beside the history of blocks. States are mapped with block numbers. With each block,
the transactions within it could potentially alter the state. Hence, we can interpret this term:
"state at block $n$": it means the state, given all the blocks from genesis up to $n$ being
executed.

First, it is important to acknowledge that maintaining the state seems optional, and it is indeed
the case. In principle, a node can decide not to maintain the state, and whenever a state value
needs to be looked up at a block $n$, all the blocks from genesis up to $n$ need to be re-executed.
This is indeed inefficient. On the contrary, maintaining a copy of the state for \textit{all} the
blocks also soon becomes a storage bottleneck. In practice, many chains adopt a middle-ground, in
which normal nodes store only the state associated with the last $k$ blocks.

Without getting into too all the details, we continue with a problem statement: in such a database,
it is very expensive for two nodes to compare their state views with one another. In essence, they
would have to compare \textit{each and every} key-value pair individually. To be able to use this
comparison more efficiently, blockchains use a data structure called a Merkle
tree\footnote{Sometimes referred to as "Trie" as well.} \cite{merkleDigitalSignatureBased1988}. A
Merkle tree\footnote{Named after Ralph Merkle, who also contributed to the foundation of
cryptography in \cite{merkleSecureCommunicationsInsecure1978}.} is a tree in which all leaf nodes
contain some data, and all non-leaf nodes contain the hash of their child nodes.

There are numerous ways to abstract a key-value database with a Merkle tree. For example, one could
hash the keys in the database to get a fixed size, base 16, string. Then, each value will be stored
at a radix-16 tree leaf, which can be traversed by this base 16 hash string.

\figuremacro{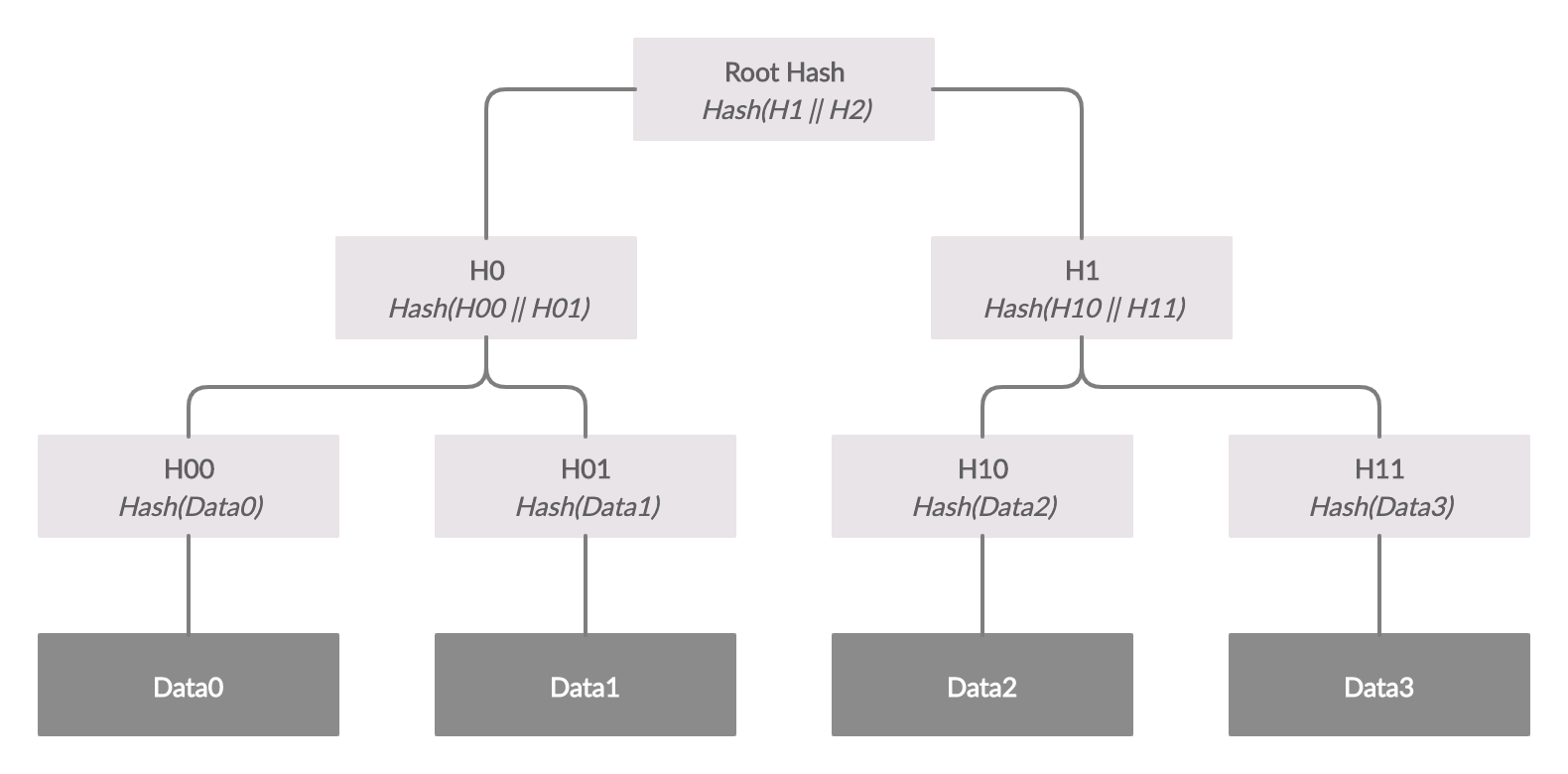}
	{Merkle Tree} {A binary Merkel Tree. The root hash contains a digest of all the 4 data nodes.}

\nomenclature{State root}{The hash of the entire state, when represented as a Merkle tree}
In such a data structure, we can clearly see that the root of the Merkle tree has a very important
property: \textit{it is the fingerprint of the \textbf{entire} data}. This piece of data is very
important in blockchains, and is usually referred to as \textbf{state root}. In essence, if two
nodes compute their individual state roots and compare them, this comparison would confidently show
if they have the same state or not. This is very similar to how the existence of the parent hash in
each block ensures that all nodes have the same chain of blocks: changing only a bit in a previous
block, or a state value in this case, will cause the hashes to no longer match. Similarly, changing
only one value in the entire key-value database will cause the state roots to mismatch.

Recalling the definition of author and validator from \ref{def:auhtor_validator}, we can now
elaborate more on what a validator exactly does. A validator node, upon receiving a block, should
check that the block's author is valid (for example check the proof of work puzzle), and then it
re-executes all the transactions in the block, to compute a new state root. Finally, this state root
is compared with the state root that the block author proposed in the block, and if they match, the
block is valid.

We can now summarize all the common data that are usually present in a block's header:

\begin{itemize}
	\item Block number: A numeric representation of the block count, also known as blockchain
	\textit{height}.
	\item Parent hash: This is the signature of the blockchain prefix.
	\item State root: It is common for a block to also name the state root that should be computed,
	if the transactions inside the block body are executed on top of the aforementioned parent hash
	block's state.
\end{itemize}

Our definition of the basic concepts of blockchain almost ends here. In the next sections, we
briefly explain concepts that are more relevant to the implementation of a blockchain, than to the
protocol itself.

\subsubsection{Runtime} \label{chap_bg:subsec:runtime}

\nomenclature{Runtime}{The portion of the blockchain's code that executes transactions}

We coin the term \textit{Runtime} as the piece of logic in the blockchain that is responsible for
\textit{updating the state}. To be more specific, the runtime of any blockchain can be simplified as
a function that takes a transaction as input, has access to read the state, and (as output)
generates a set of new key-value pairs that need to be updated in the state (or the runtime itself
can update the state directly, depending on the design of the system). This abstraction will be
further used in chapter \ref{chap:approach}.

\begin{definition} Generic Runtime. $runtime = fn(transaction) \rightarrow state$
\end{definition}

\subsubsection{Transaction Queue} \label{chap_bg:subsec:tx_queue}

By defining the transaction queue, we distinguish between transactions that are \textit{included} in
any block, and those that are not. As mentioned, a blockchain node might constantly receive
transactions, either directly from end-users, or from other nodes, as their role in some sort of
gossip protocol. These transactions are all \textit{pending}, and their existence does \textit{not}
imply anything about the state of the blockchain. Only when, by some means of consensus, everyone
agrees to append a block to the chain, the transactions within that block are included in the chain.
Thus, a transaction can be categorized as either \textit{included} and \textit{pending}.

The transaction queue is the place where all the \textit{pending} transactions live in. Its
implementation details are outside the scope of this work, and depend on the needs of the particular
chain. Nonetheless, we highlight the fact that the transaction queue is a component that sits next
to the block authoring process. Once a node wants to author a block (or it just tries to do so, in
cases such as Bitcoin, where some Proof of Work puzzle needs to be solved first), it will use the
transactions that it has received and have been stored in the transaction queue as a source of block
building.

\begin{remark}
	The transaction queue is also sometimes called transaction \textit{pool}. We prefer the term
	queue in this work because later on, we depend on the fact that a queue implies order while a
	pool implies no order.
\end{remark}

\subsubsection{Transaction Validation} \label{chap_bg:subsec:validation}

Usually, a transaction needs to pass some bare minimum checks to even be included in the queue, not
to mention being included in the canonical chain. Usually, checks that are mandatory, persistent,
and rather cheap to compute can happen right when a transaction is being inserted in the queue. For
example, the signature of a transaction must always be valid, and its validity status persists over
time. In other words, if the signature is correct, it will \textit{stay} correct over time. On the
contrary, state-dependent checks usually need to be performed when a transaction is being
\textit{included}, not when it is being inserted into the queue. The reason for this is subtle, yet
very important. If a transaction is asserting to transfer some tokens from Alice to Bob, the
state-dependent check is to make sure Alice has enough tokens. In principle, it is wrong to check
Alice's account balance at the time of inserting the transaction into the queue, since we \textit{do
not know} when this transaction is going to be \textit{included}. What matters is that \textit{at
the block in which this transaction is being included}, Alice must have enough tokens.

That being said, an implementation could optimize the read from the state in some particular way to
allow more checks to happen in the transaction queue layer (one of which is explained in the next
section, \ref{chap_bg:subsec:nonce}). Although, it should be noted that transactions in the queue
are not yet \textit{accountable}, since they are not executed. In other words, a user does not pay
any fees to have their transaction live in the queue. But, they do pay to have their transaction
included in the chain. Therefore, if the queue spends too much time on validation, this can easily
turn into a \textbf{Denial of Service} attack (DoS).

\subsubsection{Account Nonce} \label{chap_bg:subsec:nonce}

\nomenclature{Account Nonce}{An integer linked to each account and incremented per transaction. Used to prevent replay attacks}
We mentioned that signatures allow transactions to be signed only by an entity that owns a private
key associated with the account. This allows anyone to verify a transaction that claims to spend
some funds from an account. Nonetheless, given that the block history is public, this pattern is
vulnerable to \textit{replay attacks}. A replay attack is an attack in which a malicious user will
submit some (potentially signed) data twice. In the case of a blockchain, Eve can simply look up a
transaction that transfers some money out of Alice's account, and re-submit it back to the chain
numerous times. This is an entirely valid operation by itself, since the transaction that Eve is
submitting again indeed does contain a valid signature from Alice's private key.

To solve this, blockchains that rely on state usually introduce the concept of nonce: a counter that
is associated with each account in state, initially set to zero for every potential account. A
transaction is only valid if, in its signed payload, it provides the nonce value associated with the
origin, incremented by one. Once the transaction is included, the nonce of the account is
incremented by one. This effectively alleviates the vulnerability of replay attacks. Any transaction
that Alice signs, once submitted to the chain and upon being \textit{included}, is no longer valid
for re-submission.

\subsection{Putting it All Together: Decentralized State Transition Logic}
\label{chap_bg:subsec:decentralized_state_machine}

We close our introduction to blockchains by providing a final perspective on their nature. First, we
enumerate some of the lenses through which we have seen blockchains:

\begin{itemize}
	\item A distributed peer to peer network of nodes.
	\item A distributed database of blocks and states.
	\item A decentralized trustless transaction processing unit.
\end{itemize}

We can put all of this together into one frame by representing blockchains as \textbf{state
machines}. This concept resonates well with our notion of state as well. A blockchain is a
\textit{decentralized} state machine. It is a state machine because the state-root hash at the end
of each block is one potential state, and blocks allow for transition between states. Due to forks,
one might have to revert to a previous state. It is decentralized because there is no single entity
that can enforce transition from one state to another one. In fact, any participant can propose
transitions by authoring a block candidate, but it will only ever be considered canon if it is
agreed upon by everyone, through the consensus mechanism. Moreover, each participant stores a copy
of the state. If a single node crashes, goes offline, or decides to misbehave, the integrity of the
system is maintained, as long as there are enough honest participants.

\subsection{Disclaimer: A Note About The Context of Technology}

Before continuing with the chapter, we briefly address the issue of \textit{technology context}. So
far in this chapter, we have used simple examples from the banking world, since it is similar to
Bitcoin, which is a well-known system, and it is easy to explain. Nonetheless, a reader who may have
previously had some background knowledge with some other blockchain project $X$ might soon find some
details that we named here to be less than 100\% compatible with project $X$. Moreover, we have even
admitted throughout the text that some of our examples are not even exactly similar to Bitcoin (such
as the state model, as opposed to the UTXO model).

Such perceived incompatibilities/inaccuracies are predictable to happen, as blockchain systems are a
rapidly evolving field of science and engineering at the moment. Different projects diverge from one
another, even in radical concepts, and experiment with new patterns. Nonetheless, we make the
following assertions about our assumptions in this work:

\begin{itemize}
	\item Whenever we build up a simple example (mostly with Alice and Bob) in this work, we do not
	tie it to any particular blockchain project. Instead, these examples are to be interpreted
	completely independently, and solely based on the relevant concepts explained.
	\item In this entire work, we aim to see blockchains in the most \textit{generic} form that they
	can be seen. That is to say, we interpret blockchains exactly as we defined in
	\ref{chap_bg:subsec:decentralized_state_machine}: a decentralized state machine that can be
	transitioned through means of any form of opaque transaction. An example of this is the
	key-value state model that we used is more generic than the UTXO model\footnote{We have not
	explained UTXO in this work yet, but suffice to say that you can easily implement UTXO with
	key-value but not the other way around.}.
	\end{itemize}

To summarize, we have explained in this section only what we have deemed to be the most fundamental
concepts of blockchains, and we noted whenever a detail could potentially be
implementation-specific. This approach will persist throughout the rest of this work.

\section{Concurrency} \label{chap_bg:sec:concurrency}

In this section, we introduce relevant concepts from the field of concurrency. As mentioned, the
crux of our idea is to deploy concurrency in a blockchain system to gain throughput.

Concurrency is the ability of a software artifact to execute units of its logic (i.e. a
\textit{task}) \textbf{out-of-order}, without causing the outcome to be invalid
\cite{lamportTimeClocksOrdering1978}. If done correctly, this out-of-order execution can be mapped
to different hardware units and improve performance. The opposite viewpoint is a purely in-order
execution, namely sequential.

We link this directly to our example of interest: a node in a blockchain system has a process that
is responsible for executing blocks. By default, this process is purely sequential: all of the
transactions in the block are executed in order, namely the same order as they appear in the block.
Deploying concurrency \textit{should} allow this process to divide the transactions within the block
into a number of smaller groups. Then, these groups can be executed concurrently, without causing an
invalid outcome.

The outcome of interest, of course, is only the aforementioned \textbf{state database} after all of
the transactions of the block are applied. Specifically, as nodes in the network receive blocks and
apply them to a common state, the only acceptable outcome is for all of the nodes to reach the same
state after applying the block\footnote{Note that this is only the case of block validation (block
\textit{import}). There is also the task of block authoring, which is actually more complicated, but
irrelevant to the discussion of this section -- see \ref{def:auhtor_validator}.}. This comparison is
done by means of the state root.

\begin{definition} \label{def:valid_block} \textit{Valid Block}: A block is valid only if its
	execution is deterministic among all of the nodes in the network, and leads to the same state
	root $S^{'}$, if applied on top of a known previous state $S$.
\end{definition}

\begin{remark}
	The deterministic replayability of the transactions is in fact the property that ensures that
	the state is, in principle, optional to keep around, because it is \textit{deterministically
	reproducible} from the chain of blocks.
\end{remark}

Thus, the \textbf{need for determinism is absolute} in a blockchain's runtime environment. This is
in fact why blockchains are designed to work sequentially by default: because it is easy to ensure
determinism when all the transactions are applied sequentially.

Given the execution model of a block, we can reduce the problem to a single node's hardware. Assume
a node is attempting to execute a block, and it has the entire state loaded in memory. If a single
thread executes the transactions within the block, the outcome is deterministic by definition. On
the other hand, if multiple threads try to execute the transactions concurrently and access the
state as they move forward, the result is moot. This is because threads will have to compete over
access to the blockchain state and a typical race condition happens \cite{14:00-17:00ISOIEC9899}.
The challenge is to allow these threads to cooperate and achieve concurrency, while still
maintaining determinism. Therefore, we have translated our blockchain scenario to a typical
shared-state concurrency problem\footnote{We note that the determinism requirement is somewhat
special to our use case and some systems might not require it.}. In such a setup, multiple threads
are competing for access to some shared data (the state), and the runtime environment needs to
resolve the race conditions between the threads.

In the next sections, we present practical ways to use \textit{concurrency over a shared state},
while still generating valid results. Generally, mechanisms that provide more control over the
behavior of concurrent programs are referred to as  \textit{concurrency control}. Among those, we
are essentially looking for those that will allow a valid block to be authored and re-executed, as
defined in \ref{def:valid_block}, most notably deterministically.

\subsection{Locking: Pessimistic Concurrency Control} \label{chap_bg:subsec:lock}

Locks are a common and intuitive mechanism\footnote{Sadly, as we will see, the most intuitive way is
not always the easiest to use.} for concurrency control. A lock is an abstract marker applied to a
shared object (or, generally, to any memory region) to limit the number of threads that can access
it at the same time. The idea of using locks in database systems that want to achieve higher
throughput by using multiple threads goes back many decades ago
\cite{kedemControllingConcurrencyUsing1979, morrisPerformanceAnalysisLocking1985}, among other
fields.

The simplest form of a lock will not distinguish between reads and writes and the only operations on
it are \texttt{acquire} and \texttt{release}. To access the data protected by the lock, first, the
lock itself needs to be acquired. Once a thread acquires a lock, no other threads can: they have to
wait for it. Once the holding thread is done with the lock (more accurately, done with the data
protected by the lock), they release the lock. Upon being released, the lock is acquired by one of
the waiting threads, and the process restarts. This process can easily ensure that some data is
never accessed by multiple threads at the same time. The processor usually ensures that these
primitive operations (\texttt{acquire} and \texttt{release}) are done atomically between threads.
Such locks that do not distinguish between reading and writing are called a \texttt{Mutex}, short
for mutual exclusion \cite{guerraouiLockUnlockThat2019}.

A more elaborate variant of Mutex is a read-write lock (RW-lock). Such locks leverage the fact that
multiple reads from the same data are almost always harmless, and should be allowed - hence the read
and write distinction. In an RW-lock, at any given time, there can be only one writer, but multiple
concurrent readers are allowed.

\begin{remark}
	We use the Rust programming language for the implementation of this work. Rust provides some of
	the finest compile-time memory safety guarantees among all programming languages
	\cite{jungRustBeltSecuringFoundations2017}. To achieve this, Rust references (i.e., addresses to
	memory locations) have the exact same aliasing rule: multiple \textit{immutable} references to a
	data can co-exit in a scope, but \textit{only one} \textit{mutable} reference is allowed
	\cite{weissOxideEssenceRust2020}.
\end{remark}

Locks are easy to understand, but notoriously hard to use \textit{correctly}. A programmer needs to
think about every single critical memory access, and acquire and release locks from different
threads to prevent wrong outcomes. Even worse, immature use of locks often leads the programs to
deadlock, i.e., reach a  state in which all threads are infinitely waiting for a lock acquired (and
never released) by another thread. These issues are common programming errors, but they remain very
hard to detect and resolve \cite{herlihyArtMultiprocessorProgramming2012}.

Moreover, locks are a \textit{pessimistic} mechanism for concurrency control: they assume that if
two threads want to acquire the same (write) lock, their logic \textit{will} cause a conflict to
happen. Based on the granularity of the lock and the internal logic of each thread, a conflict might
not be the case all the time. This is exactly what the next section will address.

\subsection{Transactional Memory: Optimistic Concurrency Control} \label{chap_bg:subsec:stm}

Transactional memory is the opposite of locking when it comes to waiting. In locking, the threads
often need to wait for one another. If a thread is writing, then all the readers and writers need to
wait. This is based on the assumption that mutual acquirement of locks will always lead to
conflicts, so it needs to be prevented in any case. Transactional memory takes the opposite
approach, and assumes that mutual data accesses will \textit{not} conflict by default. In other
words, threads do not need to wait for one another, thus transactional memory is coined "lock-free"
or sometimes "wait-free" \cite{knightArchitectureMostlyFunctional1986}.

In the context of transactional memory, a thread's execution is divided into smaller pieces of logic
called transactions. A transaction attempts to apply one or more updates to some data, without
waiting, and then \textit{commits} the changes. Before a commit, the changes by any transaction are
not visible to any other transaction. Once a commit is about to happen, the runtime must check if
these changes are conflicting or not, based on the previous commits. If committed successfully, the
changes are then visible to all other transactions. Else, none of the changes become visible, the
transaction \textit{aborts}, and the changes are reverted. The great advantage of this model is that
if two transactions access the same memory region, but in a non-conflicting way\footnote{Textbook
example: access two different keys of a concurrent hash map.}, then there is a lot less waiting. In
essence, there is \textit{no} waiting in the execution of transactions, at the cost of some runtime
overhead when they want to commit.

Transactional memory can exist either via specialized hardware or simulated in the software,
referred to as \textit{software} transactional memory, or STM
\cite{hammondTransactionalMemoryCoherence2004}. If implemented in software, transactional memory
does incur a runtime overhead. Nonetheless, its programming interface is much easier and less
error-prone than that of locks, because the programmer does not need to manually acquire and release
locks. \cite{herlihyTransactionalMemoryArchitectural1993}.

Transactional memory is likely to lessen the waiting time and conflicts. Nonetheless, it still
allows threads to operate over the same data structure. This implies complications about commits,
aborts, and coherence\footnote{The question of which changes from a local thread's transaction
become visible to other threads and when, and under which conditions. We have barely touched this
issue, which in itself deserves a thesis to be fully understood.}. A radically different approach is
to try and prevent these complications from the get-go, by disallowing \textit{shared} data to
exist; this mechanism is described in the next section. But first, we briefly note a common trait of
locking and transactional memory.

\subsubsection*{Note About Determinism}

It is very important to note that both locking and transactional memory are non-deterministic. This
means that executing the same workload multiple times may or may not lead to the same output. It is
easy to demonstrate why: first, consider locking. Imagine two threads will soon attempt to compete
for a lock, and each will try and write a different value to a protected memory address. Based on
the fairness rules of the underlying operating system, either of them could be the first one. While
the output of the program in both cases is \textit{correct}, it is \textit{not deterministic}.

The same can be said about transactional memory: if two transactions have both altered the same data
in a conflicting way, one of them is doomed to fail upon trying to commit, and it is just a matter
of which do it first. Again, both outputs are correct, yet the program is not deterministic.

\begin{remark}
	A reader can, at this point, link our use of words "correct" and "determinism" to
	\ref{chap_bg:subsec:consensus_authorship} and block authoring specifically. From the perspective
	of a block author, it does not matter what the outcome of a block is (i.e. which transactions
	are within it, which succeed, and which fail). All such blocks are probably \textit{correct}.
	The first and foremost importance is for all validators to \textit{deterministically} come to
	the same state root, once having received the block later. This is in stark contrast with the
	non-determinism nature of locking.
\end{remark}

\subsection{Sharing vs. Communicating}\label{chap_bg:subsec:sharing_communication}

There is a great quote from the documentation of the Go programming language, which eloquently
explains the point of this section: "Don't communicate by sharing memory; share memory by
communicating." \cite{ShareMemoryCommunicating}. This introduces a radical new approach to
concurrency, in which threads are either stateless or pure functions
\cite{MostlyAdequateMostlyadequateguide}, where their state is private. All synchronization is then
achieved by the means of message passing. Like this, threads do not need to share any common state
or data. If threads need to manipulate the same data, they can send references to the data to one
another. In many cases, this pattern is advantageous compared to locking, both in terms of
concurrency degree and the programming ease.

Nonetheless, we know that our use-case exactly needs some executor threads to have a shared state
while executing the blockchain transactions. Therefore, we do not directly apply the message-passing
paradigm, but we use it as inspiration, and take the possibility of message passing into account. We
revisit this possibility in chapter \ref{chap:approach}.

\subsection{Static Analysis}

As mentioned in \ref{chap_bg:subsec:stm}, transactional memory attempts to reduce the waiting time
by assuming that conflicts are rare. This could bring about two downsides: reverts, in case a
conflict happens, and the general runtime overhead that the system needs to tolerate (for all the
extra machinery needed for transactional memory\footnote{Which is even more if it is being emulated
in the software.}). An interesting approach to counter these limitations is static analysis.
\textit{Static} refers to an action done at compile-time, contrary to runtime. The goal of static
analysis is to somehow improve the concurrency degree by leveraging only compile-time information.
In the case of transactional memory, this could be achieved by using static analysis to predict and
reduce aborts\cite{diasEfficientCorrectTransactional}. Similarly, other studies have tried to use
static analysis to improve the usage of locking by automatically inserting the lock commands into
the program's source code at compile time \cite{cheremInferringLocksAtomic2007}. This can greatly
ease the user experience of programmers using locks, and reduce the chance of human errors to
emerge. Similar to message passing, we take inspiration from the concept of static analysis in our
design later in chapter \ref{chap:approach}.

\section{Recap: Splicing Concurrency and Blockchain}

\ref{chap_bg:sec:blockchains} and \ref{chap_bg:sec:concurrency} cover the background knowledge
needed for the rest of this work. To recap, we formulated blockchains as decentralized state
machines that can be updated by the means of transactions. This state machine maps to a key-value
state database. Attempting to deploy concurrency on blockchains is, therefore, very similar to
shared state concurrency, plus the stark requirement of determinism. Shared state concurrency
attempts to resolve the problem of multiple threads accessing the same underlying memory. We
mentioned both optimistic and pessimistic concurrency, yet have not yet mentioned which one we use.
This is yet to come, in the next chapter.

\chapter{A Novel Approaches Toward Concurrency in Blockchains} \label{chap:approach}

\begin{chapquote}{Official Go Programming Language Documentation.}
    Don't share memory to communicate, communicate to share memory.
\end{chapquote}

In this chapter, we build up all the details and arguments needed to introduce our approach toward
concurrency within blockchains. We start with an interlude, enumerating different ways to improve
blockchains' throughput from an end-to-end perspective, and highlight concurrency as our method of
choice.

\section{Prelude: Speeding up a Blockchain - An Out-of-The-Box Overview}
\label{chap_approach:sec:ways_to_speedup}

Blockchains can be seen, in a very broad way, as \textit{decentralized state machines that
transition by means of transactions}. The throughput of a blockchain network, measured in
transactions per second, is a function of numerous components, and can be analyzed from different
points of view. While in this work we focus mainly on one aspect, it is helpful to enumerate
different viewpoints and see how each of them affects the overall throughput \footnote{This
categorization is by no means exhaustive. We are naming only a handful.}.

\subsection{Consensus and Block Authoring}
We discussed how the consensus protocol provides the means of ensuring that all nodes have a
persistent view of the state (see Section \ref{chap_bg:subsec:consensus_authorship}), and it can
heavily contribute to the throughput of the system. Take, for example, two common consensus
protocols: Proof of \textbf{Work} and Proof of \textbf{Stake}. They use the computation power
(\textit{work}) and an amount of bonded tokens (\textit{stake}), respectively, as their guarantees
that an entity has \textit{authority} to perform some operation, such as authoring a block. It is
important to note that each of these consensus protocols has \textit{inherently} different
throughput characteristics \cite{meneghettiSurveyEfficientParallelization2019}. Proof of work, as
the name suggests, requires the author to prove their legitimacy by providing proof that they have
solved a particular hashing puzzle. This is slow by nature, and wastes a lot of computation power on
each node that wants to produce blocks, which in turn has a negative impact on the frequency of
blocks, which directly impacts the transaction throughput. Improving the throughput of Proof of Work
requires the network to agree on an easier puzzle, that can, in turn, make the system less secure
\cite{gervaisSecurityPerformanceProof2016} (further details of which are outside the scope of this
work).

On the contrary, Proof of Stake does not need puzzle solving, which is beneficial in terms of
computation resources. Moreover, since the chance of any node being the author is determined by
their stake\footnote{Using some hypothetical election algorithm which is irrelevant to this work.},
more frequent blocks do not impact the security of the chain as much as Proof of Work does.
Recently, we are seeing blockchains turning to verifiable random functions
\cite{dodisVerifiableRandomFunction2005} for block authoring, and deploying a traditional byzantine
fault tolerance voting scheme on top of it to ensure finality
\cite{buterinCasperFriendlyFinality2019, stewartPosterGRANDPAFinality2019}. This further
\textit{decouples block production and finality}, allowing production to proceed faster and with
even less drag from the rest of the consensus system.

All in all, one general approach towards increasing the throughput of a blockchain is to
\textit{re-think the consensus and block authoring mechanisms} that dictate when blocks are added to
the chain - specifically, at which frequency. It is crucially important to note that any approach in
this domain falls somewhere in the spectrum of centralized-decentralized, where most often
approaches that are more centralized are more capable of delivering better throughput, yet they may
not have some of the security and immutability guarantees of a blockchain. A prime example of this
was mentioned in table \ref{table:blockchain_types} of chapter
\ref{chap_bg:subsec:blockchain_types}, where private blockchains are named as being always the
fastest.

\subsubsection{Sharding}
An interesting consensus-related optimization that is gaining a lot of relevance in recent years is
a technique called \textit{sharding}, borrowed from the databases field. Shards are slices of data
(in the database jargon) that are maintained on different nodes. In a blockchain system, shards
refer to sub-chains that are maintained by sub-networks of the whole system. In essence, instead of
keeping \textit{all} the nodes in the entire system synchronized at \textit{all} times, sharded
blockchains consist of \textit{multiple} smaller networks, each maintaining their own cannon chain.
Albeit, most of the time these sub-chains all have the same prefix and only differ in the most
recent blocks. At fixed intervals, sub-networks come to an agreement and synchronize their shards
with one another. In some sense, sharding allows smaller sub-networks to progress
faster\cite{forestierBlockcliqueScalingBlockchains2019, al-bassamChainspaceShardedSmart2017,
shreyDiPETransFrameworkDistributed2019}.

\subsection{Chain Topology}

Another approach to improve the throughput is changing the nature of the chain itself. A classic
blockchain is theoretically limited due to its shape: a chain has only \textit{one} head, thus only
one new block can be added at each point in time. This property brings extra security, and makes the
chain state easier to reason about (i.e. there is only one canonical chain). A radical approach is
to question this property and allow different blocks (or individual transactions) to be created at
the same time. Consequently, this approach turns a blockchain from a literal \textit{chain of
blocks} into a \textit{graph of transactions} \cite{pervezComparativeAnalysisDAGBased2018}. Most
often, such technologies are referred to as Directed Acyclic Graphs (DAG) solutions. A prominent
example of this is the IOTA project\cite{mIOTANextGenerationBlock2018}.

Allowing the chain to grow from different heads (i.e. seeing it as a graph) allows true parallelism
\textit{at the block layer}, effectively increasing the throughput. Nonetheless, the security of
such approaches is still an active area of research, and achieving decentralization with such loose
authoring constraints has proven to be challenging \cite{sompolinskySPECTREFastScalable2016}.

Altering chain topology brings even more radical changes to the original idea of blockchain. While
being very promising for some fields such as massively large user applications (i.e. "Internet of
Things", micro-payments), we do not consider DAGs in this work. We choose to adhere to the
definitions provided in chapter \ref{chap:background} as our baseline of what a blockchain is.

\subsection{Deploying Concurrency over Transaction Processing}
\label{chap_approach:subsec:out_of_box_concurrency}

Finally, we can focus on the transaction-processing view of the blockchain, and try and deploy
concurrency on top of it, leaving the other aspects, such as consensus, unchanged and, more
importantly, \textit{generic}. This is very important, as it allows our approach (to be explained
further in this chapter) to be deployed on many chains, because it is independent of any
chain-specific detail. Any chain will eventually come to a point where it must execute some
transactions, be it in the form of a chain, or a DAG, with any consensus. Thus, concurrency is a
viable solution as long as the notions of transactions and blocks exist.

Our work specifically focuses on this aspect of blockchain systems, and proposes a novel approach to
achieve concurrency within each block's execution, both in the authoring phase and in the validation
phase, thereby increasing the throughput.

\subsection{Summary} \label{chap_bg:subsec:summary_speedup}

Each of these methods brings about an improvement in the throughput (in terms of transactions
processed per second) of the system in their own unique way. The question of \textit{which one will
be the dominant one} is too specific to a particular chain's assumptions, and outside the scope of
what we are trying to tackle in this work. As an example, for some consortium chains where consensus
is less of an issue, altering the consensus is likely to result in a significant throughput gain,
without using any of the other methods such as sharding and concurrency.

Instead, we acknowledge that each of these approaches has its own merit and could be useful in a
certain scenario. Therefore, we put our focus on concurrency for the rest of this work. We see concurrency as a universal improvement that can be applied to any chain. Moreover, It is worth
noting that having optimal hardware utilization (to reduce costs) is an important factor in the
blockchain industry, as many chains are run by people who are making a profit out of running
validators and miners.

Finally, by seeing these broad options, we can clarify our usage of the word "throughput". One might
notice that the first two options mentioned in this section (consensus, chain topology) can increase
the throughput at the \textit{block} level: more blocks can be added, thus more transaction
throughput. This is in contrast to what concurrency can do. The concurrency explained in
\ref{chap_approach:subsec:out_of_box_concurrency} is the matter of what happens \textit{within}
\textit{\textbf{a}} block. Henceforth, by throughput, we mean throughput of transactions that are
being executed within a (\textit{single}) block. Similarly, by concurrent, we mean concurrent within
the transactions of a (\textit{single}) block, not concurrency between the blocks themselves.

\section{Concurrency within Block Production and Validation} \label{chap_approach:sec:concurrency}

In this section, we explain in detail how a concurrent blockchain should function. Most notably, we
define how a concurrent author and a concurrent validator differ from their sequential counterparts
by defining their requirements. Note that these requirements are mandatory and \textit{any} approach
toward concurrency in blockchains must respect them, as long as it has a common definition of a
blockchain as we named in chapter \ref{chap:background}. As a reader might expect based on previous
explanations, all of them boil down to one radical property: \textbf{determinism}.

To recap, the block author is the elected entity that proposes a new block consisting of
transactions. The block author must have already executed these transactions in some
protocol-specific order (e.g., sequentially), and noted the correct \textbf{state root} of the block
in its header. This block is then propagated over the network. All other nodes validate this block
and, if they all come to the same state root, they append it to their local chain. An author that
successfully creates a block gets rewarded for their work by the system.

\subsection{Concurrent Author} \label{chap_appraoch:subsec:concurrent_author}

We begin in a chronologically sensible way, with block authoring. Before anything interesting can
happen in a blockchain, someone has to author and propose a new block. Else, no state transition
happens.

A concurrent author has access to a pool\footnote{Also referred to as queue in
\ref{chap_bg:subsec:tx_queue}, and sometimes called \textit{mempool} in the industry.} of
transactions that have been received over the network, most often via the gossip protocol. From a
consensus point of view, it is absolutely irrelevant to ensure all nodes have the same transactions
in their local pool. In other words, from a consensus point of view, there is no consensus in the
transaction pool layer. All that matters is that any node, once chosen to be the author, has a pool
from which it can choose transactions. Then, the author has a \textit{limited} amount of time to
prepare the block and propagate it.

A number of ambiguities arise here. We dismiss them all in the following enumeration to be able to
only focus on the concurrency aspect.

\begin{itemize}
    \item Typically, the author needs some way to \textit{prefer} a subset of the transactions pool,
    as most often all of it cannot be fit into the block. For this work, we leave this detail
    generic and assume that each author has first filtered out its \textit{pool} into a new
    \textit{queue} of transactions (noting that the former is \textit{unordered} and the latter is
    \textit{ordered}) that she prefers to include in the block. In reality, a common strategy here
    is to prioritize the transactions that will pay off the most fee, as this will benefit the block
    author.
    \item The fact that the authoring time is limited is the whole reason why the author is
    incentivized to use concurrency: the more transaction that can be fitted into the block in a
    limited amount of time, the more the sum of transaction fees, thus more reward for the author.
    \item A block must have some chain-specific \textit{resource} (e.g. computation, state
    read/write, byte length of transactions) limit. For simplicity, we assume that each block can
    fit a fixed \textit{maximum} number of transactions, but this limit is so high that the
    bottleneck is not the transaction limit itself, but rather the amount of \textit{time} that that
    author has to prepare the block, bolstering the importance of high throughput. In reality, some
    blockchains have adopted a similar approach (i.e., a cap on \textit{number} of transactions), or
    have limited the size of the block. Complex chains that support arbitrary code execution even go
    further and limit the \textit{computation} cost of the transactions - see, for example,
    Ethereum's gas metering\cite{perezBrokenMetreAttacking2020}.
\end{itemize}

Having all these parameters fixed, we can focus on the block building part, namely executing each
transaction and placing it in the block.

A \textbf{sequential author} would simply execute all the transactions one by one (in some order of
preference) up until the time limit, and calculate the new state root. These transactions are then
structured as a block. Concatenated with a header that notes the state root, the block is ready to
be propagated. The created block is an \textit{ordered} container for transactions, therefore it can
be trivially re-executed deterministically by validators, as long as everyone does it sequentially.

A \textbf{concurrent author's} goal is to execute these transactions in a concurrent way, hoping to
fit \textit{more} of them in the same limited \textit{time}, while still allowing the validators to
come to the same state root. This is challenging, because, most often, concurrency is
non-deterministic. Therefore, the author is expected to piggy-back some auxiliary information to its
block, to allow validators to execute it deterministically. Maintaining determinism is the first and
foremost criterion of the concurrent author.

Moreover, the second criterion is a net positive gain in throughput. The concurrent author prefers
to be able to execute more transactions within the fixed time frame that she has for authoring, for
she will then be rewarded with more transaction fees.

\subsection{Concurrent Validator} \label{chap_appraoch:subsec:concurrent_validator}

A validator's role is simpler in both the sequential and concurrent fashion. Recall that a block is
an ordered container for transactions. Then, the sequential validator has a trivial role: re-execute
the transactions sequentially and compare state roots. The concurrent validator, however, is likely
to need to do more.

More specifically, the concurrent validator knows that a concurrent author must have executed all or
some of the transactions within the block concurrently. Therefore, conflicting transactions must
have \textbf{preceded} one another in \textit{some} way. The goal of the concurrent validator is to
reproduce the \textit{same precedence} in an efficient manner, and thus arrive at the same state
root.

For example, assume the author uses a simple Mutex to perform concurrency. In this case, some
transactions inevitably have to wait for other transactions that accessed the same Mutex earlier.
This leads to an \textit{implicit} precedence between conflicting transactions. The author needs to
somehow transfer the precedence information to the validator, and the validator must respect this
order to arrive at the same state root.

\section{Existing Approaches}\label{chap_approach:existing}

In this section, we look at some of the already existing approaches toward concurrency in
blockchains. These approaches are essentially a practical resemblance of what was explained in the
previous section. While doing so, we denote their deficiencies and build upon them to introduce our
approach.

\subsection*{Concurrency Control}

Every tool that we named for concurrency in chapter \ref{chap:background} can essentially be used in
blockchains as well, yet each takes a specific toll on the system in order to be feasible. All of
these approaches fall within the category of \textbf{concurrency control}. We begin with the
simplest one: locking.

A locking approach would divide the transactions into multiple threads\footnote{A 1:1 relation
between threads and transactions is also possible, given that the programming language supports
green threads.}. Each transaction within the thread, when attempting to access any key in the state
(Recall from \ref{chap_bg:subsec:kvdb} that the state is a key-value database), has to acquire a
lock for it. Once acquired, the transaction can access the key. This process is not deterministic.
Therefore, the runtime needs to keep track of which locks were requested by which thread, and the
\textit{order} in which they were granted. This information builds, in essence, a dependency graph.
This dependency graph needs to be sent to the validators as well. The validators parse the
dependency graph and, based on that information, spawn the required number of threads, and
distribute the transactions among them. \cite{dickersonSmartLocksAddingConcurrency2017} is among the
earliest works on concurrency within a blockchain, and adopts such an approach.

The details of generating the dependency graph with minimum size, encoding it in the block in an
efficient way, and parsing it in the validator are being highly simplified here. These steps are
critical, as they are the main overheads of this approach. The size of this graph needs to be small,
as it needs to be added to the block and increases the network overhead. Moreover, the overhead of
this extra processing must be worthwhile for the author, as otherwise, it would be in contrast to
the whole objective of deploying concurrency. There are some works that only focus on the
"dependency graph generating and processing" aspect of the process. They assume some means exist
through which the read and write set of each transaction can be computed (i.e., by monitoring the
lock requests that each thread sends at runtime). On top of this, they provide efficient ways to
build the dependency graph, and use it at the validator's end \cite{EnablingConcurrencySmart2018}.

The next step of this progression is to utilize transactional memory. This line of research follows
the same pattern. More recent works use software transactional memory (STM) to reduce the waiting
time and conflict rates. Similar to the locking approach, the runtime needs to keep track of the
dependencies and build a graph that encodes this information. Different flavours of STMs are used
and compared in this line of research, such as Read-Write STMs, Single-Version Object-based STMs,
and Multi-Version Object-based STMs
\cite{anjanaEfficientConcurrentExecution2019,anjanaEfficientFrameworkOptimistic2019}. Nonetheless,
the underlying procedure stays the same: some means of concurrency control to handle conflicts,
track dependency, and use it to encode precedence, then re-create the same precedence in the
validator.

\subsection*{Concurrency Avoidance}

\nomenclature{Concurrency Avoidance}{A scheme in contrast to concurrency control in which no synchronization is used between the threads}

Next, we name an out-of-the-box work that takes a rather different path. Many studies in the
blockchain literature use datasets from the database industry as their reference. Such datasets
might have unrealistically high rates of contention. \cite{saraphEmpiricalStudySpeculative2019} is
an empirical study that tries to determine the conflict rates within Ethereum transactions, a
\textit{live}, and arguably well-adopted network. While doing so, the study demonstrates a
different, wait-free approach. In the concurrent simulator of this work, all transactions are
executed in parallel, with the assumption that they do not conflict. If a conflict happens and a
transaction aborts, it is discarded, and re-executed again at a later phase, sequentially. This
essentially clusters transactions into two groups: concurrent and sequential. All of the concurrent
transactions are guaranteed not to conflict. The sequential transactions do not matter as they are
executed sequentially. Aside from their findings of the conflict rates in different periods of time
in Ethereum, they also report speedups in \textit{some} cases, not being too shy of the speedup
amounts reported by \cite{dickersonSmartLocksAddingConcurrency2017}, which uses locking.

This is an inspiring finding, implying that perhaps complicated concurrency control might not be
needed after all for many of the transactions in \textit{some} period of time, based on the
contention of the transactions. In some sense, this work adopts a technique that we coin as
\textbf{concurrency avoidance}, instead of \textbf{concurrency control}. As a consequence, the
system need not deal with conflicts in any way, because they are rejected and dealt with separately
in the sequential phase.

\subsection*{Static Analysis}

Finally, we note that there has also been \textit{some} work on pure static analysis in the field of
blockchains, yet all of those that we have found require fundamental changes to the programmable
language of the target chain. For example, \cite{bartolettiTrueConcurrentModel2019} provides an
extension to the Ethereum's smart contract language, Solidity, that allows it to be executed in a
truly concurrent manner (by essentially limiting the features of the language). Similarly, RChain is
an industrial example of a chain that has a programming model that is fundamentally
concurrent\cite{darrylRCast21Currency2019}, namely
pi-calculus\cite{turnerPolymorphicPiCalculusTheory1996}. Such approaches are also inspiring, yet we
prefer devising an approach which does \textit{not} need to alter such fundamental assumptions about
the programming model, in favor of easy adoption and outreach.

The aforementioned 3 broad approaches (concurrency control, concurrency avoidance, static analysis)
are essentially the taxonomy that we have found to answer the first research question, namely the
different ways to utilize concurrency within a blockchain. We close this section with a remark about
the scope of the referenced works in this chapter

\begin{remark}
    Most of the surveyed related work name their work as approaches toward concurrency for
    \textbf{smart contracts}. At this point, it would be helpful to clarify that. The details of
    smart contract chains are well beyond the scope of this work. But, it is worth noting that a a
    smart-contract chain is a \textit{fixed} chain that has a fixed state transition logic, and a
    part of that logic is to store codes (smart contracts) and execute them upon being dispatched.
    Moreover, since Ethereum is the prominent smart contract chain, all of these works present
    themselves with simulators that can hypothetically be implemented in the Ethereum node. To the
    contrary, we do not limit ourselves to smart contracts or any specific chain in this work;
    instead, we build upon the idea that the future of blockchains will not be a \textit{single}
    chain (chain maximalism), but rather an abundance of domain-specific chains interoperating with
    one another. To achieve this, one needs to think in the context of a framework for building
    blockchains, not a particular blockchain per se.
\end{remark}

\section{Our Approach} \label{chap_desgin:sec:our_approach}

In this section, we describe our approach towards concurrency in a blockchain runtime, both in the
authoring phase and in the validation phase. First, we begin by drawing a conclusion from the
surveyed studies in \ref{chap_approach:existing}.

\subsection{Core Idea: Finding a New Balance}

We begin by pointing out that all of the mentioned works, regardless of their outcome, produce a
sizable amount of overhead. We think these overheads are preventable and argue that perhaps a
different approach can prevent them altogether.

Both locking and transactional memory results in a sizeable overhead while authoring. This is mostly
hidden in some sort of runtime overhead, for example, the need to keep track of locking order, and
consequently to parse it into a dependency graph. Moreover, this dependency graph inevitably
increases the block size, because the dependency graph needs to be propagated to all other nodes.
Finally, the validator also needs to tolerate the overhead of parsing the dependency graph and
making informed, potentially complicated decisions based upon it. These are all overheads compared
to the basic sequential model. In essence, we express skepticism toward these complex runtime
machinery to deal with conflicts, and record precedence.

On the other hand, the pure \textit{concurrency avoidance} model is likely to fail under any
workload with some non-negligible degree of contention, because it basically falls back to the
sequential model where most transactions are aborted and moved to the sequential model. Results from
\cite{saraphEmpiricalStudySpeculative2019} show the same trend.

On the contrary, we aim to minimize these overheads by finding a new balance between the
"concurrency control" and "concurrency avoidance" models. Moreover, we find a new balance between
"runtime" and "static" as well. While \textit{some} runtime apparatus is needed to orchestrate the
execution and prevent chaos, tracking all dependencies is likely to be too much. Similarly, while a
purely static approach toward concurrency is a radical change to the programmable language of a
chain, we claim that \textit{some} static information could nudge the runtime to the right path.

\subsection{Key Ideas: Almost Wait-Free, Delegation, and Pseudo-Static}

Our approach is based on three key pivotal ideas, explained in the next sections.

\subsubsection{Almost Wait-Free: "Taintable" State}\label{chapt_approach:subsubsec:taintable_state}

\nomenclature{Taintable State}{Special HashMap-like data structure that assigns a taint to each key}
We have already seen that locking is a common primitive to achieve shared-state concurrency. In our
approach, we relax this primitive such that any access to a shared state by a thread \textbf{does
not incur long waiting times}, but instead, it might \textit{immediately} fail. To do so, we link
each key in the state database with a \textbf{taint} value. If a key has never been accessed before,
it is untainted. Once it is accessed by any thread (regardless of the type of operation being read
or write), it is tainted by the identifier of that accessor thread. Henceforth, any access to this
key by any other thread fails, returning the identifier of the original tainter (aka,
\textit{owner}) of the key. As we shall see in the implementation (see \ref{chap:impl}), this
approach is \textit{almost} wait-free, meaning that threads almost always proceed immediately with
any state operation. Indeed, a thread can always freely access keys that it has already tainted
before.

\subsubsection{Not Control, Nor Avoidance: Concurrency Delegation}

If a thread, in the process of executing a transaction, tries to access a tainted state key, it
forwards this transaction to the owner of that key. By doing so, a thread basically
\textit{delegates} the task of executing a transaction concurrently to another thread because it
cannot meet the state-access requirements of the transactions itself; based on the available error
information, the recipient (i.e., the owner of the failing state key) is more likely to be capable
of doing that. This is the middle ground between \textit{concurrency avoidance} and
\textit{concurrency control}, which we have coined \textit{concurrency delegation}.

Compared to concurrency control, threads in the concurrency delegation model do not try to resolve
contention in any sophisticated way. Instead, they simply delegate (aka, \textit{forward}) the
transactions to whomever they think \textit{might} be able to execute them successfully. There is no
waiting involved, and no record is kept about access precedence.

On the other hand, compared with concurrency avoidance, concurrency delegation does not
automatically assume that just because a transaction's state operation has failed, the transaction
cannot be executed in any concurrent way. Instead, the transaction might be executable by another
thread, which is known to be the owner of the problematic state key. Therefore, instead of
immediately being discarded (and potentially executed sequentially at the end), each transaction is
given a second chance to succeed in a concurrent fashion.

Finally, if a transaction is already forwarded and still cannot be executed, (only) then it is
forwarded to a sequential queue to be dealt with later. We name such transactions \textbf{orphans}.
This implies that each transaction is at most forwarded twice: once to another potential recipient,
and next, when needed, to be declared as an orphan.

Why is "a failing transaction that has already been forwarded once" considered an orphan? We present
an example to make this clearer: assume thread $T2$ receives a forwarded transaction from $T1$.
Based on the protocol of delegation, we know that this transaction must have at least one key which
is owned by $T2$, because $T1$ failed to access it and thus forwarded it to $T2$. Moreover, if the
transaction still fails, it means that it has at least one other key that is owned by some other
thread $T3$. This implies that $T2$ is not able to execute this transaction, and any further
recipient of this transaction, including $T3$, will also fail to execute it because the transaction
has at least one key owned by $T2$.

Therefore, no transaction ever needs to be forwarded more than once. A transaction can be forwarded
at most once, and thereafter it is considered to be an orphan.

\subsubsection{Pseudo-Static}

We predict that concurrency delegation works well when threads need to seldom forward transactions
to one another. In the delegation model, the transactions need to be \textit{initially} distributed
between the threads in some educated and effective way to minimize forwarding. This is where we turn
to \textit{pseudo-static heuristics} in the form of a hint. We use the term "\textit{pseudo-static}"
because we do not mean information that is necessarily available at compile-time, but is rather
known \textit{before} a particular transaction is executed. In other words, the information is
\textit{static with respect to the lifetime of a transaction} and can be inferred without the need
to \textit{execute} the transaction, but rather by just \textit{inspecting} it.

Our approach is based upon these key ideas: taintable state, concurrency delegation, and
pseudo-static hints. In the next section, we connect these ideas and depict how they work together
in the form of a unified algorithm.

\subsection{Baseline Algorithm} \label{subsec:baseline_alg}

We first look into the baseline algorithm, which essentially combines the concurrency delegation and
the taintable state, without any static heuristics. We consider a single master thread (henceforth
called "master"), and $4$\footnote{Needless to say, all of the arguments in this section are
applicable to virtually any number of threads.} worker threads (henceforth called "workers"), each
having the ability to send messages over channels to one another. The master has access to a
potentially unbounded queue of (\textit{ordered}\footnote{Recall that each node has an unordered
\textit{pool} of transactions, from which they chose an ordered \textit{queue} based on some
arbitrary preference.}, ready to execute) transactions. Moreover, it has an (initially) empty queue
of orphan transactions, to which workers can forward transactions. Additionally, each worker has a
local queue, to which the master and other workers can send transactions. The master and all workers
share a reference to the same state database, $S$, which follows the logic of a taintable state as
explained in \ref{chapt_approach:subsubsec:taintable_state}.

\begin{remark}
    As we will see, it is crucial to remember that both the transaction queue and the block are
    \textbf{ordered} containers for transactions. In other words, they both act like an ordered
    \texttt{list}/\texttt{array}/\texttt{vector} of transactions. The order of transactions in the
    queue will end up being used to order the transaction in the final block as well.
\end{remark}

The master's (simplified) execution logic during \textbf{authoring} is as follows:

\begin{enumerate}
    \item \textbf{Distribution phase}: The master starts distributing transactions between workers
    by some arbitrary function $F$. In essence, $F$ is a $fn(transaction) \rightarrow identifier$,
    meaning that, for each transaction, it outputs one thread identifier. Once the distribution is
    done, each transaction in the queue is \textit{tagged} by the identifier of one worker thread.
    The distribution phase ends with the master sending each transaction to its corresponding
    worker's local queue.

    \item \textbf{Collection phase}: The master then waits for reports from all workers, indicating
    that they are done with executing all of the transactions that they have received earlier.
    During this phase, threads might forward transactions to one another, and might forward
    transactions back to the master, if deemed to be orphan, exactly as explained in the concurrency
    delegation model. Both of these events are reported to the master and the tag of each
    transaction might change in the initial queue. Once termination is detected by the master
    thread, a message is sent to all workers to shut them down.

    \item \textbf{Orphan phase}: Once all worker threads are done, the master executes any
    transactions that it has received in its orphan queue. At this point, the master thread is sure
    that there are no other active threads in the system, thus accessing $S$ without worrying about
    the taint is safe. The transactions are executed sequentially on top of $S$, and their tag is
    changed to a special identifier for orphan transactions.
\end{enumerate}

Then, the master is ready to finalize the block. By this point, each transaction is either tagged to
be orphan, or with the identifier of one of the $4$ worker threads. In essence, we have clustered
transactions into $4 + 1$ \textit{ordered} groups. The validator who receives this block respects
this clustering and executes transactions with the same tag in the same thread. In the first $4$
groups, the transactions within a group might conflict with one another (e.g. attempt to write to
the same key), but they are ordered and are known the be executed by the same thread, thus
deterministic. The transactions in the last group, namely the orphan group, are executed
sequentially and in isolation, thus deterministic.

The worker side of the processing (with slight simplifications) is as follows:

\begin{enumerate}
    \item \textbf{Depleting local queue}: Having received a number of transactions from the master
    (after the "\textit{Distribution phase}"), each worker then tries to deplete its local queue.
    For each transaction $Tx$, the logic is as follows:

    If $Tx$ is executed successfully, nothing is done or reported. This is because the master is
    already \textit{assuming} that $Tx$ is executed by the current worker, therefore nothing needs
    to be reported. If $Tx$ fails due to a taint error, it is forward to the owner of the state key
    that caused the failure. Note that at this point we know that $Tx$ has not been forwarded
    before, because it is being retrieved from the initial local queue.

    At the end of this phase, the worker sends an overall report to the master, noting how many of
    the transactions it could execute successfully, and how many ended up being forwarded. This data
    is then used at the master to detect termination.

    \item \textbf{Termination phase}: Once done with their local queue, the workers listen for two
    types of messages, namely \textit{termination} or \textit{forwarded} transactions from other
    threads. Termination is the message from the master to shut down the worker. Forwarded
    transactions are those that another worker is \textit{delegating}/\textit{forwarding} to the
    current worker because of a taint error. The forwarded transaction is then executed locally and,
    if it is successful, the result is reported to the master. If the execution fails again due to a
    taint error, then the transaction is forwarded to the master as an orphan.

    Note that in this case, reporting is vital, because a worker is ending up executing a
    transaction and the master is \textit{not} aware of it, because the worker who finally executes
    the transaction is not the same as assigned in the "Distribution phase" of the master. This
    reporting is also needed for the termination detection of this phase.
\end{enumerate}

\subsubsection{Analysis and Comments on the Baseline Algorithm}
\label{chap_approach:subsec:comment_baseline}

A number of noteworthy remarks exist on the baseline algorithm:

\textbf{Termination Detection}. We intentionally did not describe how the master detects the
termination of the collection phase, because, if we wanted to describe it, we needed further
information from the worker's logic. Recall that the master knows how many transactions it has
initially distributed between all the workers. Moreover, from the reports sent by the worker at the
end of "Depleting local queue" phase, it knows how many of them executed in their \textit{designated
thread}, and how many of them ended up being \textit{forwarded}. Also, recall that each forwarded
transaction, upon being executed successfully, is reported to the master. Similarly, each forwarded
transaction that fails is also reported to the master (by being forwarded to the orphan queue
residing in the master thread). Thus, the master can safely assert that \textit{Termination is
achieved once the sum of "all locally-executed transactions at workers", "forwarded and successfully
executed transactions" and "orphan transactions" is equal to the initial count of transactions in
the queue}. At this point, the termination message is created and broadcasted to all workers.

\begin{definition}
    \textbf{Termination of the master thread's collection phase. }

    Assume $N$ initial transaction. Each worker thread, upon finishing the "deplete local queue"
    phase, reports back $\{ r_{1}, r_{2}, r_{3}, r_{4} \}$ respectively, indicating the number of
    transactions that each executed locally. Given $D$ as the number of reports of transactions
    being forwarded, and $O$ as the size of the orphan queue, termination of the collection phase is
    achieved iff

    \begin{equation}
        \sum_{t = 1}^{4} r_{t} + D + O == N
    \end{equation}
\end{definition}

\textbf{Maintaining Order}: Aside from termination detection, it is also vital for determinism that
the master takes action upon the report of a transaction being forwarded. This is because
\textbf{once the tag of a transaction changes, it is likely that its order must also change within
the queue}. For example, if a transaction, initially assigned to $T_{1}$ is known to be forwarded
and executed by $T_{2}$, it is important to re-order it in the initial transaction queue such that
it is placed \textit{after} all the transactions initially assigned to $T_{2}$. This is because, in
reality, $T_{2}$ \textit{first} executed all of its designated transactions and \textit{then}
executes any forwarded transactions. Recall that the queue is an ordered container for transactions
and its order will eventually end up building the order of the transaction in the block.

\textbf{Orphan Transactions}. An orphan transaction is a transaction that has already been forwarded
and still fails to execute at its current host thread, due to a taint error. We now represent this
from a different perspective. In our concurrency delegation scheme, threads race to access state
keys and, upon successful access, they taint them. Any transaction has a number of state keys that
it needs to access in order to be processed. \textit{An orphan transaction is one that has state
keys being tainted by at least two \textbf{different} threads}. For example, assume a transaction
needs to access keys $K_{1}$ and $K_{2}$. Assume thread $T_{0}$ is executing this transaction. If
$K_{1}$ is already tainted by a $T_{1}$ and $K_{2}$ by a $T_{2}$, then this transaction will
inevitably end up being in the orphan queue. The transaction is first forwarded from $T_{0}$ to
$T_{1}$, where it can successfully access $K_{1}$, but still fails to access $K_{2}$ and thus
orphaned\footnote{An interesting optimization can be applied on top of this logic, which is
explained further in \ref{chap_impl:sec:opt_orph}.}.

\textbf{Minimal Overhead}. Our approach incurs minimal overhead to the block. In fact, the only
additional data needed is one identifier attached to each transaction, indicating which thread must
execute it (and the special case thereof, orphan transaction), namely the \textit{tag}. This can be
as small as a single byte per transaction, which is negligible. Note that transactions within the
block still maintain partial order: the transactions of a particular tag are sorted within their
tag. Only the relative order of transactions from different tags is lost, which is not significant,
because they are guaranteed by the author to not conflict.

\textbf{Validation}. We can now consider validation as well. As expected, due to the minimal
overhead and the simplicity of the baseline algorithm, the validation logic is fairly simple. Each
block is received with all of its transactions having a tag. The ones tagged to be orphans are set
aside for later execution. Then, one worker thread is spawned per tag, and transactions are assigned
to threads based on their tags, and in the same relative order. The workers can then execute
concurrently, \textit{without the need for any concurrency control}, because they effectively know
that all contentious transactions already have the same tags, and thus are ordered
\textit{sequentially} \textit{within that tag/thread}. In essence, the validation is fully
\textit{parallelizable} and does not need any synchronization. Once all threads are finished, the
orphan transactions are executed sequentially and validation comes to an end.

\subsubsection{Determinism in Concurrency Delegation}

Finally, we must address the most important requirement: \textbf{Determinism}. We prove determinism
by showing that the validation and authoring both have the exact same execution environment, and the
transactions are executed in the exact same order in both phases. In more detail, both the author
and the validator spawn the same number of threads, and all of the transactions executed by any
thread during authoring is re-executed by a single thread in the validation phase as well.

First, consider all transactions that are executed in their initially designated worker, based on
the aforementioned distributor function $F$. All of these transactions are assigned a tag, and have
some partial order within the tag. By design, they are also placed in their designated worker
thread's queue with the same order, and thus executed in the same order. Consequently, they are, yet
again, placed in the final block with the same order within the tag. Therefore, the worker thread in
the authoring phase and the worker thread in the validation phase will execute \textit{exactly the
same} transactions, in the exact same order.

Next, consider the transactions that ended up being forwarded. These transactions are executed
\textit{after} the designated transactions of their final host thread. The master thread,
responsible for building the final block, has to note this change and ensure that this partial order
is maintained in the final block. The first step for the master is to change the tag of this
forwarded transaction to the tag of its newly designated thread. Then, to ensure that the order
\textit{within the tag} is maintained, the master simply places this transaction at the end of the
queue. This ensures that this transaction will be placed in the final block in such a way that it is
executed after all the transactions that have the same tag. Then, we can apply the same logic as the
previous paragraph and assert that the order of execution of all transactions with a specific tag
stays the same within one thread in both authoring and validation, thus deterministic.

Last but not least, the orphan transactions also need the same property to be executed
deterministically: maintaining order. The master needs to make sure that all the transactions that
are tagged as \texttt{Orphan} within the block have the same order as they were executed.

Given the deterministic execution of all types of transactions in our concurrent system, we conclude
that our approach is fully deterministic, with minimal additional effort . Indeed, the only subtlety
that needs to be taken care of is the re-ordering of a transaction in the queue (and consequently
the block) when it is \textit{successfully} forwarded.

\subsection{Applying Pseudo-Static Heuristics} \label{subsec:applying_static_hints}

The previous section is a complete description of a concurrent system that is deterministic and can
be deployed as-is, without any further requirements. Nonetheless, one can argue that this system
might not be efficient in the throughput gain that it can deliver. This is because we said nothing
about the distribution function of the master, namely $F$. While we keep this distribution function
generic in this entire work, we make one important claim about it: using static information of the
transaction can be \textit{very} beneficial, and is key to high performance.

Static information is anything that can be known about a transaction, before it hits the runtime and
gets executed. In other words, we are interested in any information that can be inferred from the
transaction, \textit{without} needing to \textit{execute} it. A clear example of such information is
the origin of the transaction (i.e., the sender account). Because of public-key cryptography usage,
all transactions carry a signature and the origin account as a part of their payload. Therefore,
using the origin as static information is permitted, because it is known even \textit{before} the
transaction is executed. Similar reasoning applies to the arguments of the transaction as well.
These are information that is encoded in the payload of the transaction, and the runtime of the
master thread (before starting authoring) can effectively use them to optimize $F$. On the contrary,
consider the return value of the transaction. This is a piece of information that is not considered
static with respect to the transaction, because it can only be known by executing the transaction.

\begin{remark}
    Given the above paragraph, we denote that our definition of static is different from the term
    which is usually referred to as compile-time information. Therefore, we used the term
    pseudo-static in some places to delineate the difference.
\end{remark}

We use this pseudo-static information to our benefit, by proposing a static annotation to be added
by the programmer to each transaction. Recall that the underlying state of the blockchain runtime is
a key-value database, so each state access is linked to a key. Moreover, if $F$ can know the list of
keys accessed by each transaction, it could create a perfect distribution where no transaction is
ever forwarded or orphaned, because no thread ever reaches a taint error while accessing the state.
Of course, things are not simple. In practice, it is impossible to know the execution path of a
complex transaction without executing it\footnote{An interested reader can refer to the "Halting
problem" for a formal representation of this issue \cite{burkholderHaltingProblem1987}.}, therefore
knowing the exact state keys that it must access is impossible.

The important point is to remember that the annotation can still be reasonably accurate without the
need for executing the transaction. This annotation can state, in a best-effort manner, which state
keys are \textit{likely} to be accessed by this transaction, based on, for example, one of the
common execution paths of the transaction. Listing \ref{lst:static_pseudo} shows an abstract example
of these static \textbf{annotations}.

\begin{lstlisting}[caption={Example of Static Hints}\label{lst:static_pseudo}]
#[access = (origin.state_key(), arg1.state_key())] fn transaction(origin, arg1, arg2) {read(origin);
if condition {// more probable branch! read(arg1)} else {// less probable branch! read(arg2)}}
\end{lstlisting}

We observe a transaction, embodied as a function named \texttt{transaction}, which has 3 arguments:
the origin and two auxiliary ones. Furthermore, we observe a macro\footnote{This style refers to the
Rust programming language, but is applicable to any other compiled language as well.} that is
providing the state keys that \textit{might} be accessed by this transaction. In this case, the
second argument is not relevant, and seemingly only some state key of the origin and the
\texttt{arg1} might be accessed.

This macro syntax is just one example - its specification is irrelevant at this point. What matters
is that the transaction can provide the runtime with some easy-to-compute, pseudo-static information
about the \textit{state access} of the transaction (therefore we called the macro \texttt{\#access}
in listing \ref{lst:static_pseudo}). The runtime can then effectively use this information in the
transaction distribution phase to come up with a better distribution that leads to less contention,
and, consequently, to less forwarding and fewer orphans.

\begin{remark}
    One might question: why is the system limited by the halting problem and cannot pre-execute all
    transactions in the first place? This is partially answered in the transaction validation
    section (see \ref{chap_bg:subsec:validation}). The more detailed reason is that this static
    hinting must fit into the transaction validation pipeline, because the transactions for which we
    are keen to know the access keys are not yet included in any block, thus not accountable. In
    short, executing all transactions in-order pool is an easy Denial Of Service attack vector,
    especially in chains that allow arbitrary code execution by the user.
\end{remark}

\subsection{System Architecture}

In this section, we bring \ref{subsec:baseline_alg} and \ref{subsec:applying_static_hints} into one
picture, and explain the entire system architecture. The all-in-one system architecture is presented
in Figure \ref{figure:figures/design.png}. In the following paragraph, we explain all its
components, as indicated by the labels in the figure, finally connecting all the dots of our
concurrent system as we move forward.

\figuremacro{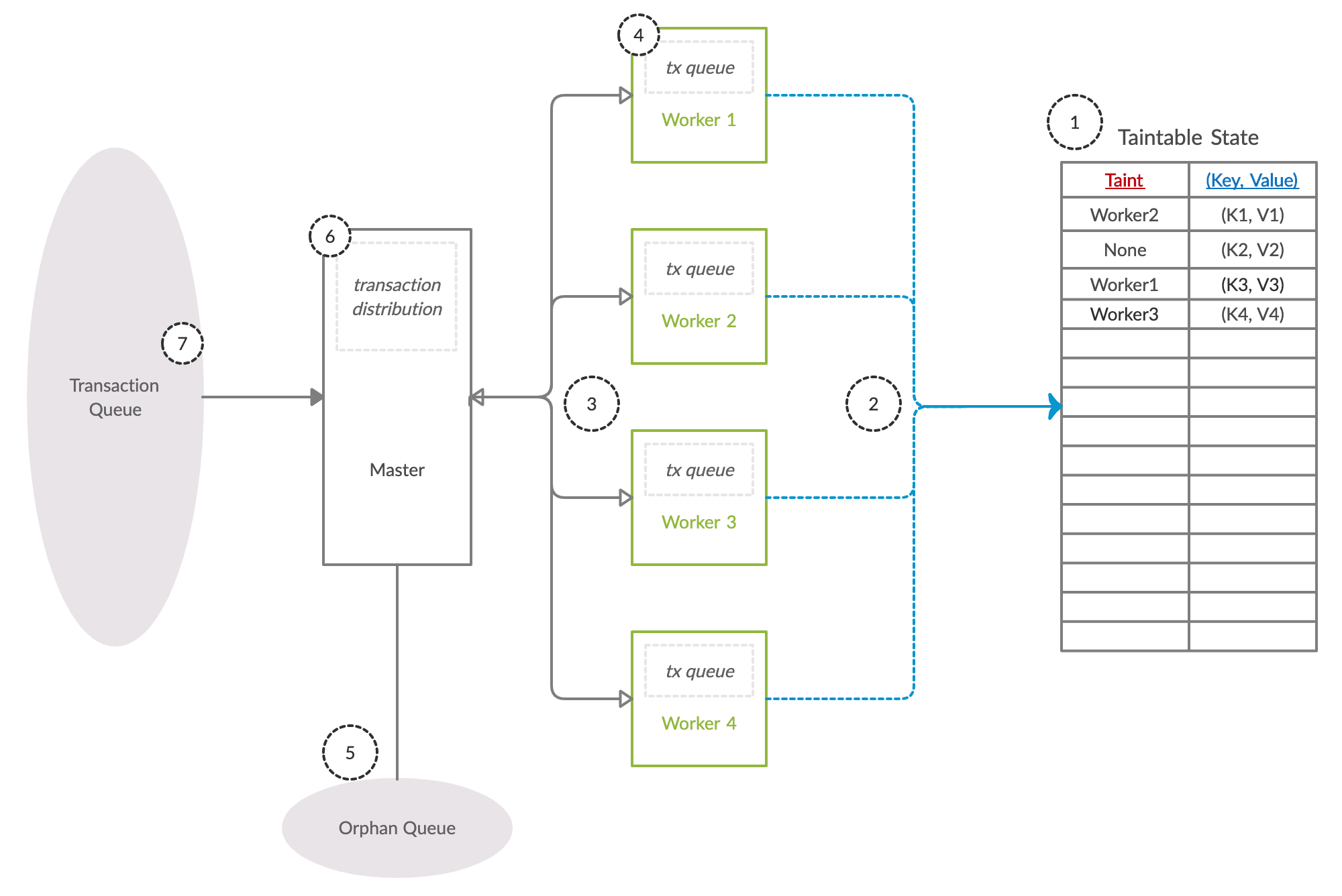}{Overall System Architecture.}


\begin{itemize}
    \item \textbf{1} shows the taintable state. We can see that each cell is not merely a key-value
    pair, but rather a \texttt{(taint, key, value)} triplet. Some keys are not tainted
    (\texttt{None}), while some are already tainted by a thread because they have been accessed by
    it. Access to each of the keys by a thread succeeds if the key is untainted, or is tainted by
    the accessor thread. Else, an error is returned, notifying the identifier of the thread that
    owns the key.
    \item \textbf{2} emphasizes the need of all threads to access the state; ideally, this access is
    wait-free. Any access to a key in the state either \textit{immediately} succeeds, in which case
    the key is tainted. Else, it is \textit{immediately} rejected because of a taint error.
    \item \textbf{3} shows the communication channels between the master and workers. These channels
    allow all of the threads to communicate by means of sending messages to one another. Most often
    these messages are simply a transaction to be forwarded. It is worth noting that worker threads
    can also communicate in the same manner.
    \item \textbf{4} indicates the local queue of each thread. This is where the transactions that
    the master thread designates to each thread live until they are executed. Needless to say, this
    queue is also ordered.
    \item \textbf{5} identifies the orphan queue, where the master thread maintains an ordered queue
    of transactions that are essentially rejected by the workers and need to be executed in a second
    sequential phase. Once all the workers are done accessing the state, the master thread
    exclusively starts using the state (essentially ignoring all the taint values) and executes all
    the orphan transactions sequentially.
    \item \textbf{6} importantly depicts the transaction distributor component of the master thread.
    This component is invoked prior to the process of authoring to tag all the transactions that are
    ready to be executed in the transaction queue. This effectively determines which thread gets to
    execute which transactions.
    \item \textbf{7} is the transaction queue, where an ordered subset of the transaction pool is
    verified and awaiting to be delegated to worker threads.
\end{itemize}

With this overall blueprint of the system's architecture, we conclude the design of SonicChain and
move forward to the next chapter, where we cover some of the implementation details of our
prototype.

\chapter{Implementation} \label{chap:impl}

\begin{chapquote}{u/goofbe on reddit}
    Rust is like a futuristic laser gun with an almost AI-like foot detector that turns the safety
    on when it recognizes your foot.
\end{chapquote}

In this chapter, we bring the system architecture mentioned at the end of the chapter
\ref{chap:approach} closer to a running prototype. This chapter is by no means extensive since there
are many many implementation details that could be worth noting. Nonetheless, in favor of brevity,
we minimize the details to only those that:

\begin{itemize}
    \item Are important with regards to the evaluation of the system.
    \item Impose a particular practical challenge that we find interesting.
\end{itemize}

The entire source code of the prototype is available as free and open-source software
\cite{paimaniKianenigmaSubSonic2020}.

For the implementation, we use the Rust programming language
\cite{klabnikRustProgrammingLanguage2019}. Being backed by Mozilla, Rust has a lot to offer in the
domain of system programming and low-level applications, such as a blockchain runtime. Rust is a
unique language, among the few rare ones in which one can claim that the learning curve is indeed
steep, as mastering it is \textit{not} just a matter of learning the new syntax. One of the reasons
for this learning curve is Rust's compile-time memory management system, which means all allocations
and de-allocations are inferred and checked \textit{at compile time}. This ensures that the program
is memory-safe, even in a multi-threaded context\footnote{The Rust community often uses the term
"Fearless Concurrency" for this combination -- memory safety and concurrency
\cite{FearlessConcurrencyRust}}, whilst having no garbage collector at runtime. Rust delivers, in
some sense, the performance of C, combined with the abstractions and safety features of Java or
C\#\cite{jungRustBeltSecuringFoundations2017}.

\begin{remark}
    We assume some basic Rust knowledge in the rest of this chapter. An interested reader without
    any Rust experience can also follow, yet they are likely to have to look up some types and
    concepts in the documentation.
\end{remark}

Lastly, it is worth mentioning that our choice is not merely out of interest. In the last few years,
Rust has been heavily invested in, by big blockchains companies and in their research
\cite{RustBlockchain}.

\section{The Rust Standard Library}

First, we explain some of the primitive types available in Rust's standard library that we use
in our implementation. Note that these types are merely \textit{our choice} for this implementation.
Although similar data types from other libraries (or \textit{crates}, in the Rust jargon) are also
acceptable, we prefer to limit ourselves to the standard library and remain dependency-free in our
proof-of-concept implementation.

\subsection{State: HashMap and Locks} \label{chap:impl:subsec:state_and_concurrent_map}

For the taintable state, we need a data type that is \textit{similar} to a typical concurrent
\texttt{HashMap}\cite{barnatFastDynamicallySizedConcurrent2015}, yet has slight differences. Rust
does not provide a concurrent HashMap in the standard library, so the way to go is to
implement our own custom data structure. To implement this, we use a \texttt{HashMap} and a
\texttt{RwLock}, both of which are provided by the standard library. The HashMap behaves just like a
typical \texttt{HashMap} in any programming language. The \texttt{RwLock} is a locking primitive
with read-write distinction, where multiple \textit{read} requests can be done at the same time,
while a \textit{write} request will block access of all-but-one requests.

Like most data types in languages that support generics, Rust's default \texttt{HashMap} is
generic over both the key and value type that it uses. The final \texttt{HashMap} is using opaque
byte arrays as both the key and value type. For the keys, we do our own hashing and concatenation to
compute the location (i.e., the final key) of any given state variable. For example, to compute the
key of where the balance of an account is stored, we compute: \texttt{"balances:balance\_of".hash()
+ accountId.hash()} and use the final byte array as the key. As for the values, to be able to store
values of different types in the same map, we encode all values to a binary format, thus, a byte
array is used.

\begin{lstlisting}[caption={Key and Value Types (with slight simplification -- see \ref{lst:state_final}). }\label{lst:maps}]
/// The key type.
type Key = Vec<u8>;
/// The value type.
type Value = Vec<u8>;

/// Final State layout.
type State = HashMap<Key, Value>;
\end{lstlisting}

Listing \ref{lst:maps} shows how the final state type is created. Namely, we alias the state to be a
hashmap with both key and values being opaque byte arrays.

\begin{remark}
    A careful reader might notice at this point that we explained how we build the \textit{key} byte array (hashing and concatenation), yet we have not yet discussed how the value byte array is built. This will be explained in \ref{chap_impl:sec:bonus}.
\end{remark}

\subsection{Threading Model}

In short, Rust's threading model is \texttt{1:1}: each thread will be mapped to an operating system
thread (which sometimes directly maps to a hardware thread). Naturally, this means that the overhead
of spawning new threads is not negligible, but there are no additional runtime overheads. This
justifies using a small number of threads and assigning a large number of transactions to each,
rather than, for example, creating one thread per transaction.

The main purpose of this decision is for Rust to remain a \textit{runtime-free} programming
language\cite{RustJourneyAsync}, meaning that there is zero-to-minimal runtime in the final binary.
The opposite of the \texttt{1:1} threading model is the \texttt{N:M} model, also known as
\textit{green threads}. Such threads are lightweight, and do not map to a hardware/operating-system
thread in any way. Instead, they are a software abstraction and are handled by a
runtime\footnote{tokio is one of the best known such runtimes in the Rust
ecosystem\cite{TokioRust}.}. Therefore, a language that wants to support green threads needs some
sizeable runtime machinery to be able to handle that.

\subsection{Communication}

For communication, we use multi-producer-single-consumer\cite{StdSyncMpsc} (\texttt{mpsc} for short)
channels - the only type available in the standard library. \ref{figure:figures/design.png} depicted
the communication medium for the threads as something similar to a bus, where all threads can
directly communicate with all other threads. In reality, the layout of the channels is a bit more
complicated.

The process of using Rust's \texttt{mpsc} channels is such that each channel has one receiver and
one producer handle, and the producer handle can be freely copied into different threads (while the
the receiver cannot be copied around -- this is ensured by Rust's compile-time checks). With this
approach, each thread has an \texttt{mpsc} channel and keeps the receiving handle to itself, while
giving a copy of the producer handle to all other threads.

Listing \ref{lst:mpsc} demonstrates this process. A producer and receiver pair is created at line
2. Further down, two threads are created, where each receives a \texttt{clone()} of the producer. The
receiver will stay in the starting thread, and it is used at the end to checks for incoming
messages.

\begin{lstlisting}[caption={How Channels Allow Communication Between Threads (with slight simplification)}\label{lst:mpsc}]
// In the local thread.
let (producer, receiver) = std::mpsc::channel();

// spawn a new thread.
std::thread::spawn(|| {
    // this scope is local to a new thread.
    let local_producer = producer.clone();
    // note the cloned handle ----^^^^^^^

    local_producer.send("thread1");
});

// spawn another new thread.
std::thread::spawn(|| {
    // this scope is local to a new thread.
    let another_local_producer = producer.clone();
    // note the cloned handle -----------^^^^^^^

    another_local_producer.send("thread2");
});

// check incoming messages.
while let Ok(msg) = receiver.rcv() {
    // do something with `msg`.
}
\end{lstlisting}

\section{Example Runtime: Balances} \label{chap_impl:sec:balances}

Next, we demonstrate an example runtime to help the readers get familiar with the context of the
implementation and how the final outcome looks like. Recall from \ref{chap_bg:subsec:runtime} that
a runtime is the core state transition logic of each chain. Our final implementation supports
multiple runtime \textit{modules} within, each having their own specific business logic. The
simplest example of such a module is a \texttt{balances} module that takes care of storing the balance
of some accounts and allows the transfer of tokens between them. We now enumerate some of the
important bits of code involved in this module.

First, there needs to be a \texttt{struct} to store the balance of a single account, which we named
\texttt{AccountBalance} - see listing \ref{lst:balance_struct} for the full definition.

\begin{lstlisting}[caption={Balance Strict}\label{lst:balance_struct}]
/// The amount of balance that a certain account.
#[derive(Debug, Clone, Default, Eq, PartialEq, Encode, Decode)]
pub struct AccountBalance {
    /// The amount that is free and allowed to be transferred out.
    free: u128,
    /// The amount that is reserved, potentially because of the balance being used in other modules.
    reserved: u128,
}
\end{lstlisting}

The state layout of these modules is simple: there needs to be one mapping, with the key being an
account identifier and the value being \texttt{AccountBalance}, as in listing
\ref{lst:balance_struct}.

\begin{lstlisting}[caption={State Layout of the Balances Module}\label{lst:balance_state}]
decl_storage_map!(
    // Auxillary name assigned to the storage struct.
    BalanceOf,
    // Auxillary name used in key hashing.
    "balance_of",
    // Key type.
    AccountId,
    // Value type.
    AccountBalance,
);
\end{lstlisting}

Note that the statement in listing \ref{lst:balance_state} is a macro (denoted by the \texttt{!}
notation), meaning that it generates a substantial amount of code at compile time. Most of this code
deals with functions for generating the key of a specific account's balance, namely the hashing and
concatenation method explained in \ref{chap:impl:subsec:state_and_concurrent_map}. To recap, the
code generated by this macro means: the final state key of an any account in the storage is computed
as: \texttt{"balances:balance\_of".hash() + account.hash()}.

Overall, the inline documentation of the listing should make it clear that: there exists a state
\textit{mapping} from \texttt{AccountId} to \texttt{AccountBalance}". We have already seen what
\texttt{AccountBalance} is, and \texttt{AccountId} is an alias for --no surprise-- a public key.

Finally, we can look at the only public transaction that can be executed in this module: a transfer
of some tokens from one account to another one, presented in listing \ref{lst:balance_tx}.

\begin{lstlisting}[caption={Transfer Transaction}\label{lst:balance_tx}]
#[access = (|origin| vec![
    <BalanceOf<R>>::key_for(origin),
    <BalanceOf<R>>::key_for(dest)
])]
fn transfer(runtime, origin, dest: AccountId, value: Balance) {
    // read the balance of the origin.
    let mut old_balance = BalanceOf::read(runtime, origin).or_forward()?;

    if let Some(remaining) = old_balance.free.checked_sub(value) {
        // origin has enough balance. Continue.
        old_balance.free = remaining;

        // new balance of the origin.
        BalanceOf::write(runtime, origin, old_balance).unwrap()

        // new balance of the destination.
        BalanceOf::mutate(runtime, dest, |old| old.free += value).or_orphan()?;

        Ok(())
    } else {
        Err(DispatchError::LogicError("Does not have enough funds."))}
    }
\end{lstlisting}

The logic of the transaction should be straightforward: (1) read the origin's balance; (2) if they
have enough free balance, update the balance of the origin and destination, (3) Else, return an
error.

The more interesting bit is in lines 1-4 of listing \ref{lst:balance_tx}, where our long
promised \texttt{access} macro is being used in action. The interpretation of the macro is basically
as follows: this transaction will (most likely\footnote{Recall that the \texttt{access} macro was
supposed to be a best-effort \textit{guess}.}) access two state keys, the balance of the origin and
the balance of the destination.

\section{Generic Distributor}

Another noteworthy detail of the implementation is the distributor component. Recall that the
distributor is responsible for tagging each transaction in the transaction queue with the identifier
of one thread. We emphasize that we leave this detail \textit{generic} in our implementation,
similar to its position in chapter \ref{chap:approach}. This means that there is no concrete
implementation of a distributor in our system. Instead, any function that satisfies a certain
requirement can be plugged in and used as the distributor. The main detail to remember is that we
\textit{prefer} the distributor to use the hints provided by the \texttt{access} macro. because (see
chapter \ref{chap:approach}) such a distributor will be a lot more effective at preventing
transaction forwarding.

Two examples of distributor implementation are as follows:

\begin{itemize}
    \item \textbf{Round Robin}: This distributor simply ignores the \texttt{access} macro and
    assigns transactions to threads one at a time, in a sequence.
    \item \textbf{Connected Components}\cite{nuutilaFindingStronglyConnected1994}: This graph
    processing algorithm is the exact opposite end of the spectrum compared to Round Robin, meaning
    that it \textit{heavily} takes the \texttt{access} macro into account.

    Specifically: all transactions provide a list of state keys that they might access during their
    execution, via the \texttt{access} macro. This distributor builds a bipartite graph of
    transactions and state keys, where each transaction has an edge to all the keys that it might
    access. Therefore, two transactions that are likely to access the same state key end up being
    \textit{connected}, because they both have an edge to that key. The connected component, as the
    name suggests, identifies these connected transactions. Every group of transactions that access
    a common set of state keys is grouped as a component. Once all components are identified, they are distributed among threads as evenly as possible\footnote{Using a simple greedy algorithm
    that we do not describe here.}. In essence, keeping a full component as a unit of work
    distribution ensures that transactions that might conflict will end up being sent to the same
    thread, effectively minimizing forwarded and orphaned transactions.
\end{itemize}

\section{Bonus: Optimizing Orphans} \label{chap_impl:sec:opt_orph}

A closer look at \ref{lst:balance_tx} reveals that the first state access error is being handled by
\texttt{.or\_forward()?} and the second one with \texttt{.or\_orphan()?}. The reason for this is
quite interesting: Recall from \ref{chap_approach:subsec:comment_baseline} that an orphan
transaction is basically one that needs access to state keys that are tainted by at least two
different threads. From this, we can realize: if a transaction successfully accesses any state key,
this means that no other thread can execute this transaction, because the key associated with that
first state access is already tainted. Therefore, we can conclude a massive simplification: Only if
the first state access of a transaction fails it will be forwarded. Any further failure is simply
an orphan right off the bat.

\section{Bonus: Taintable State} \label{chap_impl:sec:bonus}

So far, all the mentioned details of this chapter are somewhat necessary to be able to comprehend
the evaluation in chapter \ref{chap:bench_analysis}. Conversely, this last section is optional, and
explains some of the details of the Taintable state implementation. This explanation does assume
even more Rust knowledge from the reader.

Let us recap the situation in the state \texttt{HashMap}. We know that the state is basically a
\texttt{HashMap<Vec<u8>, Vec<u8>>}, and we already know how the key is constructed (hashing and
concatenation of some prefixes and values). Two questions arise:

\begin{itemize}
    \item How is the \textit{value} constructed? For example the \texttt{decl\_storage\_map!()} in
    the balances module mapped an \texttt{AccountId} to an \texttt{AccountBalance}. How exactly is
    the \texttt{AccountBalance} encoded to \texttt{Vec<u8>}?
    \item We claimed that the state is almost wait-free. How can the underlying implementation
    support this? As noted, our tainting logic is very special to our use case, and the
    implementation of a typical concurrent \texttt{HashMap} will not be a good inspiration for us,
    because it would have different waiting semantics.
\end{itemize}

The answers are actually connected and are provided together.

Recall that our initial proposal was for the state to be wait-free, meaning that any access to the
state would succeed or fail immediately. We can solidify this idea as such: \textit{only the first}
access to each key of the state \textit{might} incur some waiting for other threads; thereafter, all
operations are wait-free. It should be clear why we can obtain this property: if a key is accessed
just once, it is tainted, and, therefore, all further accesses can be immediately executed based on
that taint value. Specifically, following access from the owner thread succeeds, while a following
access from another thread fails. It is, unfortunately, impossible to allow \textit{all} accesses to
be wait-free, because a write operation to a key that had not existed before could trigger a resize
of the HashMap. Now it is clear why we titled SonicChain \textit{almost} wait-free.

Next, we discuss how this behavior is implemented. First, we explain what the actual value stored in
the state is. Listing \ref{lst:state_value} shows the data type stored in the state. Note that the
generic \texttt{struct} has separate \texttt{taint} and \texttt{data} fields. Moreover, the taint is
\textit{optional}, meaning that it may or may not exist , as denoted by \texttt{Option<\_>} in listing \ref{lst:state_value}.

\begin{lstlisting}[caption={The state value type}\label{lst:state_value}]
pub struct StateValue<V, T> {
    /// The data itself.
    data: RefCell<V>,
    /// The taint associated with the data.
    taint: Option<T>,
}
\end{lstlisting}

The final state type is defined in listing \ref{lst:generic_state_type}. Note that the key type, the value
type, and the taint type are all left out generic.

\begin{lstlisting}[caption={The final generic state type}\label{lst:generic_state_type}]
pub type StateType<K, V, T> = HashMap<K, StateValue<V, T>>;
\end{lstlisting}

Now we can re-iterate on listing \ref{lst:maps} and correct it, leading to the implementation in
listing \ref{lst:state_final}.

\begin{lstlisting}[caption={The final (\textbf{concrete}) state type}\label{lst:state_final}]
pub type Key = Vec<u8>;
pub type Value = Vec<u8>;
pub type ThreadId = u8;

pub type State = StateType<Key, Value, ThreadId>;
\end{lstlisting}

To recap, the \texttt{Key = Vec<u8>} is computed from hashing and concatenation, and the
\texttt{Value = Vec<u8>} is the binary encoding of any data type that we may store in the state -
for example \texttt{u128} for the case of balances (as seen in \ref{lst:balance_struct}).

Finally, we can demonstrate the locking procedure. To implement the state, we wrap a
\texttt{StateType} in a \texttt{RwLock} inside a new \texttt{struct}. This \texttt{struct} will then
implement appropriate methods to allow the runtime to access the state - see listing
\ref{lst:struct_taint_state}.

\begin{lstlisting}[caption={The wrapper for \texttt{StateType}}\label{lst:struct_taint_state}]
/// Public interface of a state database.
pub trait GenericState<K, V, T> {
    fn read(&self, key: &K, current: T) -> Result<V, T>;
    fn write(&self, key: &K, value: V, current: T) -> Result<(), T>;
    fn mutate(&self, key: &K, update: impl Fn(&mut V) -> (), current: T) -> Result<(), T>;
}

/// A struct that implements `GenericState`.
///
/// This implements the taintable struct. Each access will try and taint that state key. Any further
/// access from other threads will not be allowed.
///
/// This is a highly concurrent implementation. Locking is scarce.
#[derive(Debug, Default)]
pub struct TaintState<K: KeyT, V: ValueT, T: TaintT> {
    backend: RwLock<StateType<K, V, T>>,
}

impl<K, V, T> GenericState<K, V, T> for TaintState<K, V, T> {
    // implementation
}
\end{lstlisting}

The most interesting piece of code omitted in \ref{lst:struct_taint_state} is the \texttt{//
implementation} part of \texttt{GenericState} for \texttt{TaintState}: this where we define how and
when we use the \texttt{RwLock} of \texttt{backend} in \texttt{TaintState}.

Specifically, each access to the backend will first acquire a read lock (which is not blocking, and
other threads can also access it at the same time). Two possible outcomes exist:

\begin{itemize}
    \item If the key is already tainted, the state operations will fail or succeed trivially: all
    non-owner threads will receive an error with the thread identifier of the owner, and all owner
    operations will succeed. Moreover, note that the inner data in the map itself is wrapped in a
    \texttt{RefCell}, which allows interior mutability\cite{RefCellInteriorMutability}. In essence,
    this means that the \textit{owner} thread can manipulate the data only by acquiring a read lock. Our tainting logic and rules ensure that this will not lead to any race conditions.

    \item If a key is not already tainted (or non-existent in the inner \texttt{HashMap}), then the
    thread proceeds by trying to acquire a write lock. This is needed, because this operation is going to alter the \texttt{taint} field of a \texttt{StateValue}, and all other threads need to be blocked. Once the taint has been updated, the write access is immediately dropped, so that all other threads can proceed.
\end{itemize}

One final important remark is of interest. We mentioned Rust's concurrent memory safety with much
confidence in earlier sections of this chapter, claiming that it can prevent memory errors in
compile time. Nonetheless, we see that using a \texttt{RefCell}, we can alter some data, even when
shared between threads, without any synchronization. How is this possible?

The key is that Rust does allow such operations in \texttt{unsafe} mode. The "unsafe mode" of Rust,
sometimes called the "wild west" of Rust, is based on a contract between the programmer and the
compiler, where the programmer manually testifies that certain operations are memory safe, and need
not be checked by the compiler. In some sense, unsafe Rust is like a fallback to \texttt{C}, where
memory can be arbitrarily accessed, allocated, de-allocated, and such.

In our example, we have such a contract with the compiler as well. We know that:

\begin{itemize}
    \item If all threads access the \textit{taint} with a \textbf{write} lock, and
    \item If all threads access the \textit{data} after checking the taint, via a \textbf{read} lock.
\end{itemize}

All race conditions are resolved, thus no compiler checks are needed.

Indeed, this contract needs to be delivered to the compiler by a single \texttt{unsafe} statement in
our state implementation - listing \ref{lst:unsafe}. Here, \texttt{Sync} is a trait to mark data
types that are safe to be shared between threads. As the name recommends, it is a marker trait with
no functions. \texttt{RefCell} is not \texttt{Sync} by default, because it allows arbitrary interior
mutability. Because our \texttt{StateValue} contains a \texttt{RefCell}, it is also not
\texttt{Sync} be default. The statement in listing \ref{lst:unsafe}, which needs to be \texttt{unsafe}, tells
the compiler to make an exception in this case and allow \texttt{StateValue} to be used in a multi-threaded context. This allows us to wrap the \texttt{TaintState} in an \texttt{Arc} (atomic
reference counted pointer) and share it between threads.

\begin{lstlisting}[caption={Unsafe Implementation in State}\label{lst:unsafe}]
unsafe impl<V, T> Sync for StateValue<V, T> {}
\end{lstlisting}

\chapter{Benchmark and Analysis}  \label{chap:bench_analysis}

\begin{chapquote}{Michael A. Jackson}
    Rules of Optimization:
    Rule 1: Don't do it.
    Rule 2 (for experts only): Don't do it yet.
\end{chapquote}

\ref{chap_impl:sec:balances} introduced an example runtime module in our implementation, namely a
balances module that can store the balance of different accounts and initiate transfers between
them. In this chapter, we build upon this module and provide benchmarks to evaluate SonicChain, as
described in \ref{chap_desgin:sec:our_approach}. First, we begin by explaining the details of the
benchmarking environment, including the data set.

\section{The Benchmarking Environment}

All experiments are executed on a personal laptop with \texttt{2,3 GHz 8-Core Intel Core i9 CPU} and
\texttt{32 GB 2400 MHz DDR4 RAM}. We keep the machine connected to power for consistent results, and
run no additional resource-intensive software while taking measurements.

We measure the execution time of both the \textit{authoring} and \textit{validation} tasks. From the
computed time, we derive the throughput in \textit{transactions per second}. Recall that authoring
is the process of creating a block, and validating is the task of re-importing it to ensure
veracity; these tasks are performed by the \textit{author} and \textit{validator}, respectively.
Moreover, recall that in our concurrency delegation model, by the end of the authoring phase, all
transactions are tagged with the identifier of the thread that should execute them. Therefore, the
validation task is fairly simpler. Furthermore, we set the \texttt{access} macro of the transfer
transaction to point to the account balance key of the \texttt{origin} and \texttt{destination}
account, as demonstrated in listing \ref{lst:balance_sig}.

\begin{minipage}{\linewidth}
\begin{lstlisting}[caption={Signature of the Transfer and its Access Hints}\label{lst:balance_sig}]
#[access = (|origin| vec![
    <BalanceOf<R>>::key_for(origin),
    <BalanceOf<R>>::key_for(dest)
])]
fn transfer(runtime, origin, dest: AccountId, value: Balance) { /* implementation */ }
\end{lstlisting}
\end{minipage}

Then, we use this information to spawn two benchmarks, one with \textbf{connected components} and
one with a \textbf{round robin} distributor. Indeed, we also use a \textbf{sequential} version as
baseline.

The \textbf{dataset} is composed of two parts: the \textbf{initial state} and the
\textbf{transactions}.

The initial state is the state of the world \textit{before} any transactions are executed. In our
case, this maps to a number of initial accounts and an initial balance in each of them. The initial
balance can be parameterized. The second parameter is the number of transactions between the
accounts. Recall that the only transaction in our balances module is \texttt{transfer}. For example,
assuming 100 accounts and 50 transactions, each of the 50 transactions is generated by picking two
random accounts (e.g., Alice and Bob) from the entire set of 100 accounts and creating a
\texttt{transfer(Alice, Bob, Amount)} with a fixed transfer amount.

In this chapter, we only focus on a variation of this dataset that we call \textit{the millionaire's
playground}. This is because we assign a very large amount of initial balance to each account,
ensuring that it is many times larger than the transfer amount. Consequently, all transfers will
\textbf{succeed}.

Note that despite the transfer being a dead-simple transaction, the success or failure branches have
different \textbf{state access requirements}. Namely, a succeeding transaction accesses the
balance of both the origin and the destination of the transfer, while a failing one only
accesses the former. This, next to the value placed as a hint in the access macro can lead to
interesting combinations that are outside the scope of this chapter, but will be discussed
further in chapter \ref{chap:conclusion}.

We assume that there are no time limits imposed on the author of the block, and allow it to finish
executing all transactions. Of course, as we already delineated earlier, this is not how things work
in reality (see \ref{chap_bg:subsec:consensus_authorship}). Nonetheless, this model allows us to be
able to clearly see the throughput difference between the sequential execution and the concurrent
ones.

From within authoring, we do not measure the execution time of the generic distributor. A critical
reader might think this is a way for us to avoid taking the execution time of connected components
into account. We strongly assert otherwise. In most modern blockchains, authors know, well in
advance, if and when they author blocks. Therefore, it is \textit{very} sensible for them run the
distributor procedure (be it the expensive connected components, or the cheap round-robin, or
anything else) over their pool of transactions in advance, as a form of pre-processing. Even if they
do not know when they might author a block, it is rather straightforward to
\textit{periodically/regularly} run the distributor on their local pool\footnote{Lastly, we
anecdotally realized that running connected components is not a real bottleneck in graphs with
sizes in the order of a few thousand transactions, and with our experimental setup.}. For the
simpler task of validation, we measure the entire process of parsing the block and executing it.

We must also address the state storage issue. In a real-life scenario, the state database is likely
to be kept on high latency storage, and therefore access to new keys might be orders of magnitude
slower (particularly the first one, depending on the caching) than any computation. Our
implementation keeps the entire state in an in-memory \texttt{HashMap}. We acknowledge that this is
likely to be too simplistic and to compensate, we artificially insert \texttt{sleep} operations into
the read and write operations of the final state implementation.

Lastly, we assume that both the authoring and validation uses the same degree of concurrency,
meaning that everyone has the same number of worker threads, ready to receive tasks and
transactions.

\section{Benchmark Results} \label{chap_b&a:sec:results}

For the first demonstration, we fix the number of accounts and gradually create more (transfer)
transactions between them. For all 3 classes of executions (sequential, round-robin, connected
components) we generate 1000 members and increase the number of transactions from 250 to 2000. Both
the validation and authoring times are measured. All executions utilize 4 worker threads. Table
\ref{table:bench1} presents all the results in one picture.

\begin{table}[h]
\centering
\caption{Benchmarking results with the millionaire's playground data set. All times are in ms. RR and CC stand from round robin and connected components distributors, respectively. "4" refers to the number of workers.}
\label{table:bench1}
\resizebox{\textwidth}{!} {
    \begin{tabular}{l|rr|rr|rr}
    \textbf{type}    & \textbf{members} & \textbf{transactions} & \textbf{authoring (ms)} & authoring tps   & \textbf{validation (ms)} & validation tps  \\ \hline
    Sequential       & 1000             & 250                   & 1899                    & \textbf{131.65} & 1905                     & \textbf{131.23} \\
    Sequential       & 1000             & 500                   & 3639                    & \textbf{137.40} & 3797                     & \textbf{131.68} \\
    Sequential       & 1000             & 1000                  & 7548                    & \textbf{132.49} & 7508                     & \textbf{133.19} \\
    Sequential       & 1000             & 2000                  & 15053                   & \textbf{132.86} & 14969                    & \textbf{133.61} \\ \hline
    Concurrent(RR-4) & 1000             & 250                   & 898                     & \textbf{278.40} & 704                      & \textbf{355.11} \\
    Concurrent(RR-4) & 1000             & 500                   & 2161                    & \textbf{231.37} & 1789                     & \textbf{279.49} \\
    Concurrent(RR-4) & 1000             & 1000                  & 5517                    & \textbf{181.26} & 4545                     & \textbf{220.02} \\
    Concurrent(RR-4) & 1000             & 2000                  & 12788                   & \textbf{156.40} & 10797                    & \textbf{185.24} \\ \hline
    Concurrent(CC-4) & 1000             & 250                   & 625                     & \textbf{400.00} & 453                      & \textbf{551.88} \\
    Concurrent(CC-4) & 1000             & 500                   & 1210                    & \textbf{413.22} & 887                      & \textbf{563.70} \\
    Concurrent(CC-4) & 1000             & 1000                  & 9510                    & \textbf{105.15} & 7162                     & \textbf{139.63} \\
    Concurrent(CC-4) & 1000             & 2000                  & 19698                   & \textbf{101.53} & 14820                    & \textbf{134.95}
    \end{tabular}
}
\end{table}

Despite our first benchmark being a simple one, it already unravels plenty of details and
hidden traits about the system's behavior. Thus, we make the following observations:

\subsection*{Sequential}
The sequential execution is nothing special. As expected, the execution time of both tasks
increases linearly as the number of transactions increases. The throughput, as expected, stays
more or less the same.

\subsection*{Round Robin}
The behavior of the round-robin distributor degrades over time. With a small number of transactions
(e.g. 250), the throughput increase in authoring is more than 100\%. As the number of transactions
grows more toward 1000, the increase drops. At 1000 transactions, authoring is only a smidgen 20\%
more than the sequential throughput.

The reason for this behavior is of interest. Recall that round-robin will distribute the
transactions between the threads with no particular knowledge. This is expected to cause a fairly
large number of transactions to become forwarded or orphans. Our execution logs clearly demonstrate
this behavior. For example, we have the following lines from the execution log of round-robin with
250 transactions.

\begin{lstlisting}
[Worker#3] - Sending report Message { payload: AuthoringReport(42, 20), from: 5 }. From 42 executed, 42 were ok and 0 were logic error.
[Worker#1] - Sending report Message { payload: AuthoringReport(45, 18), from: 3 }. From 45 executed, 45 were ok and 0 were logic error.
[Worker#2] - Sending report Message { payload: AuthoringReport(46, 16), from: 4 }. From 46 executed, 46 were ok and 0 were logic error.
[Worker#0] - Sending report Message { payload: AuthoringReport(46, 17), from: 2 }. From 46 executed, 46 were ok and 0 were logic error.
[Master  ] - Finishing Collection phase with [179 executed][32 forwarded][39 orphaned]
\end{lstlisting}

Lines 1-4 contain the \texttt{AuthoringReport} message sent from the worker to master. The two
numeric fields of this message are the number of transactions that got executed and forwarded
respectively. As seen in the log, each worker notified the master that it failed to execute a
portion of their designated transactions. Line 5 is the aggregate information log of the master
thread once all the transactions, except the orphans, are executed. As seen in the log, from the 250
transactions, 179 were executed in their \textit{designated} thread, 32 were forwarded and executed,
and finally, 39 were orphaned. This clearly shows the consequence of round robin's \textit{blindness}
toward the \texttt{access} macro. The amount of transactions that got forwarded and orphaned
directly contributes to a reduction in throughput.

Now, we can see the equivalent log in the execution with 1000 transactions.

\begin{lstlisting}
[Master] - Finishing Collection phase with [389 executed][146 forwarded][465 orphaned]
\end{lstlisting}

With 1000 transactions, more than half of them failed to execute in their designated thread, and
ended up being forwarded or orphaned. This further demonstrates why the throughput drops from 250 to
1000 in round-robin.

As for validation, we see that the validation throughput is analogous to that of authoring in each row, but slightly better. Moreover, similar to authoring, the throughput drops as we increase the
transactions. It is very important to understand why. Recall that during validation, the validator
simply uses the tags provided by the author to know which transaction needs to be executed where.
Now, let us analyze the destiny of the imperfect transactions, namely forwarded ones and orphans.
The forwarded transactions will be treated by the validator as if they were not forwarded. These
transactions are still simply assigned to a thread, and the validator can effectively save time by
executing them concurrently, in multiple threads. On the other hand, the orphan transactions are a
\textit{loss} of throughput for \textit{both} the author and the transaction. If a transaction is
declared as an orphan, not only the author, but also the validator both lose any throughput gains for
that transaction. This clarifies the throughput trend of the validation in round-robin, and how it
drops as we increase the transactions. The underlying reason is, in fact, that as the number of
transactions grows, the number of orphans also increases in the round-robin benchmarks. Therefore, a
decline in the throughput of the validator is also expected.

\subsection*{Connected Components}
The outcome of the connected components is even more interesting. For transaction counts 250 and 500
we see much better throughput than both sequential and round-robin. This trend applies to both
authoring and validation. Nonetheless, for transaction counts 1000 and 2000 we observe the
throughput plummets down to rates lower than the sequential throughput.

The reason for this can also be explained by examining the logs. First, let us look at the same log
lines for one of the \textit{good} execution, namely 500 transactions.

\begin{lstlisting}
[Worker#2] - Sending report Message { payload: AuthoringReport(125, 0), from: 24 }. From 125 executed, 125 were ok and 0 were logic error.
[Worker#1] - Sending report Message { payload: AuthoringReport(125, 0), from: 23 }. From 125 executed, 125 were ok and 0 were logic error.
[Worker#0] - Sending report Message { payload: AuthoringReport(125, 0), from: 22 }. From 125 executed, 125 were ok and 0 were logic error.
[Worker#3] - Sending report Message { payload: AuthoringReport(125, 0), from: 25 }. From 125 executed, 125 were ok and 0 were logic error.
[Master  ] - Finishing Collection phase with [500 executed][0 forwarded][0 orphaned]
\end{lstlisting}

Interestingly, this time all worker threads executed all of their designated transactions with no
error, leaving no particular leftover work to do for the master thread. Such cases lead to slightly
less than 4-fold throughput gain in authoring and a perfect 4-fold gain in validation.

Now let us examine what goes wrong in the case of 1000 transactions.

\begin{lstlisting}
[Worker#3] - Sending report Message { payload: AuthoringReport(8, 0), from: 29 }. From 8 executed, 8 were ok and 0 were logic error.
[Worker#2] - Sending report Message { payload: AuthoringReport(8, 0), from: 28 }. From 8 executed, 8 were ok and 0 were logic error.
[Worker#1] - Sending report Message { payload: AuthoringReport(8, 0), from: 27 }. From 8 executed, 8 were ok and 0 were logic error.
[Worker#0] - Sending report Message { payload: AuthoringReport(976, 0), from: 26 }. From 976 executed, 976 were ok and 0 were logic error.
[Master  ] - Finishing Collection phase with [1000 executed][0 forwarded][0 orphaned]
\end{lstlisting}

In this interesting case, jumping into line 5 of the logs that gives an overview actually unravels
absolutely nothing: It still seems like all the transactions are executed in their designated
thread. The problem is hidden in the number of transactions that each thread received. As seen in
lines 1-4, the first 3 threads received a total of 24 transactions, while all the remaining
transactions were given to the last worker thread. The reason for this is rooted in the number of
accounts and transactions. Given 1000 accounts, generating 1000 transfers from them is likely to
create a large blob of \textit{interconnected} transactions. This is something that the connected
components cannot deal well with. The outcome is that a very large chunk of transactions will be
assigned to one giant component, consequently given to one worker thread to execute. This is,
essentially, a typical work imbalance problem that can be seen in different fields of parallel and
concurrent computing. Given this, the connected component execution, in this case, is basically
sequential, \textit{and} it has a whole lot of overhead from message passing. Therefore, the
throughput drops significantly, even below the sequential one.

As for the validation, we can use the conclusions from the round-robin section to reason about
its behavior. Recall that forwarded transactions are not an overhead for the validator, and only
orphan transactions can cause an overhead. In the case of 1000 and 2000 transactions and connected
components, the validation throughput is almost the same as that of the \textbf{sequential}
execution. This is because the distributor essentially linearizes the transactions into a sequential
group. Moreover, all of the overhead is for the author. Therefore, it is expected that the
throughput of the validator is almost the same as sequential execution.

The results indicate different traits and characteristics of the entire system as a whole, and
our two distributor components of choice. Round robin is an example of a system in which we use
concurrency but without any pseudo-static hints. On the contrary, connected components is a prime
example of scenarios where the decision of transaction distribution is entirely based upon
pseudo-static hints. While the results are in favor of the latter, we also observed that in some
niche scenarios, connected components is inflexible and therefore turns out to \textit{not} be the
best option. In the next chapter, we summarize these details into the conclusion of our work.

\chapter{Conclusion} \label{chap:conclusion}

\begin{chapquote}{Edsger Dijkstra}
    Simplicity is a prerequisite for reliability.
\end{chapquote}

We have embarked on a long journey to reach this stage of our work. To make the conclusion more
comprehensible, we first briefly recap what we have done so far, and what we have observed in the
benchmarks of chapter \ref{chap:bench_analysis}. We then enumerate our conclusive observations in
\ref{chap_conc:sec:discussion}. We then answer the research questions in \ref{chap_conc:sec:rq}.
Finally, we mention some of the future work to be done in \ref{chap_conc:sec:future_work}.

\section{Summary}

We began by making minimal, yet important assumptions about what a blockchain system should look
like, whilst explaining the details thereof in chapter \ref{chap:background}. Looking back in
hindsight, the most important assumption that we have is the key-value based \textit{state}
implementation. With the state, we can analogize a runtime executing transaction over a
\textit{state} to a thread in a multi-core CPU, trying to access \textit{memory} as it executes
code. We are then faced with a problem with a setup very similar to that of the shared state
concurrency, except the absolute need for determinism. In essence, determinism is the main
blockchain-specific challenged imposed on this problem.

We identified the de-facto way of solving this challenge in contemporary literature to be building
dependency graphs and piggy-backing them into the block. With further investigation, we decided to
not proceed further down this path: instead, we considered that a much simpler model can be
sufficient, \textit{if} we utilize all of the information available.

In chapter \ref{chap:approach} we declared "\textit{a much simpler model}" to be the
\textit{concurrency delegation} model, where conflicting transactions can be forwarded between
threads when possible, or executed sequentially, at the end of the processing "cycle", otherwise. In
essence, we remove the chance of non-determinism by allowing threads to \textit{only} execute
non-conflicting transactions. In other words: \textit{intra-thread} transactions can conflict as
much as they want, but \textit{inter-thread} transactions must not conflict at all. All transactions
that generate inter-thread conflicts are deemed to be orphan, and will be executed sequentially. The
main advantage of the delegation model is its final outcome: because of the absence of inter-thread
conflicts, the only bit of data that needs to be added to each transaction is one final identifier
about which thread should execute it. The validator is guaranteed to be able to deterministically
re-execute the block, whilst gaining potential throughput benefits.

Furthermore, we make sure we "\textit{utilize all of the information available}" by adding some
pseudo-static hints on top of each transaction. These hints need to be provided by the programmer,
and denote the state keys that are likely to be accessed by the transaction. These hints do not need
to be highly accurate. Moreover, we restricted our hints to data that is cheap to know prior to the
execution of the transaction. Basically, the dilemma is to, not solve, but rather find a way around
the halting problem\cite{burkholderHaltingProblem1987}. To do so, we acknowledge that, most often,
transactions are simple units of logic, with specific behavior given different
inputs\footnote{Blockchain transactions are expensive pieces of code to execute, because they need
to be re-executed by hundreds, if not thousands of other nodes. They are not designed to execute
complex arbitrary logic, and are unlikely to evolve in this direction in the foreseeable future.}.
Given this, it is likely to be easy for the programmer to provide somewhat accurate hints about the
behavior of a transaction.

We combined these features in a concurrent system for high-throughput transaction processing called
SonicChain. Furthermore, we provided a prototype implementation of the system, where we included the
\texttt{access} macro and the \textit{distributors} - see Chapter \ref{chap:impl}. We used this
prototype to evaluate an example application called "the millionaires' playground". We further
benchmarked the performance of this example application. The results, presented in chapter
\ref{chap:bench_analysis}, show throughput improvements for both distributor types, when executed
against a synthetic balance transfer workload.

\section{Discussion and observations} \label{chap_conc:sec:discussion}

Our work has lead to the following observations.

\textbf{The transactions are \textit{key}}. Needless to say, the type of transactions that are being
executed, and the amount of inter-transaction state dependencies between them, is the first and
foremost factor of throughput improvement. This factor trumps the underlying system as well. With a
workload that is made almost entirely from interdependent transactions, even the best of systems
probably fails to deliver high throughput. Moreover, with a workload with almost no
inter-transaction dependencies, probably most systems perform well. We showed, in our experiments,
that our system is no different in that regard. Nonetheless, the important point is that we managed
to achieve ideal throughput gains by effectively leveraging the static hints of the transactions.

\textbf{Simplicity}. Our work is, in some sense, a retaliatory demonstration against complicated
concurrency control mechanisms deployed (mostly in academia) on blockchains. Our main goal was not
to trump their results and prove that our system is better per-se. Rather, we showed that for
certain workloads (that are --reasonably speculating-- not far from reality), a much simpler system
will also be enough. Moreover, the new and simpler system is actually beneficial upon a certain
axis, for example, smaller runtime and block overhead.

\textbf{Generic Design}. This work is generic at two different levels. \textbf{First}, our
\texttt{access} macro is merely one example of how the static\footnote{As before, by static we mean
static with respect to the execution of the transaction.} data that a transaction carries can be
used. Implementations could vary, or different means of analysis can be used. The only point is to
remain aware that the goal is to be able to infer the state access requirements of transactions
\textit{\textbf{without executing them}}. \textbf{Second}, with or without the \textbf{access}
macro, we leave the decision of \textit{how to use the output of static hints} to be generic in the
form of component that we called the distributor. We do provide two such distributors, merely to
illustrate the difference in complexity, and discuss their processing cost, but many other variants
can be envisioned.

\section{Research Questions and Answers} \label{chap_conc:sec:rq}

Based on this work, the answers to our research questions (formulated in section
\ref{chap_intro:sec:resarch_q}) are as follows.

\begin{enumerate}
    \item [\textbf{RQ1}] What approaches exist to achieve concurrent execution of transactions
    within a blockchain system, and improve the throughput?

    \item [\textbf{Answer}] We identified the current usages of concurrency within blockchains to
    within one of the two categories that we depicted: concurrency control or concurrency avoidance.
    Both do result in a throughput gain. Full answer in \ref{chap_approach:existing}.

    \item [\textbf{RQ2}] How could both static analysis and runtime approaches be combined together
    to achieve a new approach with minimum overhead and measurable benefits?

    \item [\textbf{Answer}] We basically opted to reduce the runtime apparatus (coined: concurrency
    delegation) and instead compensate with pseudo-static advisory data that can complement the
    runtime to achieve an equally good result as concurrency control but with less overhead in
    validation. Full answer in \ref{chap_desgin:sec:our_approach}.

    \item [\textbf{RQ3}] How would such an approach be evaluated against and compared to others?

    \item [\textbf{Answer}] We used an empirical evaluation approach to measure the throughput gain.
    A \textit{realistically} synthesized data set was created for this purpose. We concluded that
    our approach is sufficient to achieve ideal throughput gains (given the number of threads) under
    certain circumstances, which supports our claim that a simpler system could be well enough.
\end{enumerate}

Summarizing, this work makes the following contributions:
\begin{enumerate}
    \item A new runtime model for concurrency with minimal overhead, namely concurrency Delegation
    \item A proposal to couple runtime machinery with pseudo-static information for better result.
    \item A first prototype of such a system, namely SonicChain, implemented in the Rust programming
    language.
    \item An empirical analysis of the system against realistic workloads.
\end{enumerate}

\section{Future Work} \label{chap_conc:sec:future_work}

Our discussion and observations already hint at some of the future work that we propose to be
pursued. Here, we enumerate such future work directions in more detail.

\textbf{Inaccurate hints}. We boldly mention that static hints do not need to be accurate for the
system to work properly. Of course, they should not be totally unrelated. A good starting point to
reason about hints is to look at the transaction's logic and proceed with the control flow
optimistically or pessimistically. Which flow has the most state access? Which one has the least?
The interesting follow-up question is how does the behavior of the system change when we move
between different access macros? One might recall, the transfer transaction that we used before had
a pessimistic access macro: we assumed the balance of both the \texttt{origin} and the
\texttt{destination} will be accessed. This is not always the case. If the origin does not have
enough funds, then the transaction finishes with an error, and only the balance key of the origin is
accessed (i.e. \textit{tainted}). We excluded such experiments from our work here for the sake of
brevity.

\textbf{Probabilistic Hints}. An interesting addition to the previous point is the mixing of
probability theorems with the access macro. In essence, instead of relying on the programmer to
provide the hints, the system could use benchmarks and previous data to build probabilistic models
to accurately predict the access requirements of a transaction given specific inputs. This basically
transforms the access macro from being (pseudo)\textit{static} to \textit{probabilistic}. At the
very extreme end of this spectrum, one could even see the utilization of artificial intelligence to
accurately generate the access requirements of a transaction.

\textbf{Read-Write Taint}. A comment on our work that can fundamentally question our approach is:
why is there no distinction between read and write operations? This is a fair comment, and we agree
that it is an unorthodox approach compared to the rest of the literature in concurrent computing. We
did not opt for this approach in our work because of our strong preference for simplicity. There are
many paths to follow along the line of read-write distinction. Most would make our system quite
similar to a cache coherence model, where a write operation needs to be notified to others and
potentially invalidate other previous read operations. This is already a red line for us, as we do
not want to carry the burden of rollbacks, both because of their overhead and inherent
non-determinism. A final interesting comment about this aspect would be that read-write taint is
also likely to introduce read-only \texttt{access} macro hints. We foresee that it will be very
useful for the distributor to be able to know if an access hint is read-only or not. All in all, we
foresee interesting outcomes if the path of read-write taints is pursued, but express worry about
how many complications it will add to the system, compared to the actual throughput gain. The
obvious benefit of this addition to the work would be that threads that only \textit{read} certain
data will not block other threads.

\textbf{Continuous Analysis of Real Chains}. After all, everything that we have done here is in some
sense hypothetical, because we do not use real transaction from a real chain. It is of great value
for further studies to analyze different, domain-specific (e.g. Bitcoin) and general-purpose chains
(e.g. Ethereum), and derive characteristics from their transactions (as it was already done in
\cite{saraphEmpiricalStudySpeculative2019}). For example, an easy study is to apply connected
components to different blocks of the Ethereum network and see how well they can be clustered into
disjoint components with a balanced size. Will we face the same imbalance problem as with our
experiments in chapter \ref{chap:bench_analysis}?

\textbf{Hybrid distributor.} In our work we did not explore the possibility of hybrid distributors.
This can also be a fairly simple addition to the system, making it more versatile. Even only within
the two distributors that we discussed, we could see that it was not the case that one distributor
is the best in all cases. Round robin had limited throughput gains, but it was still better than the
ideal connected components in data sets that had highly intertwined transactions. We propose
building hybrid distributors that potentially chose the best distribution function based on
particular criteria, specific to the application and/or dataset at hand.

\textbf{Nonce-aware execution}. We explained what a nonce is in \ref{chap_bg:subsec:nonce}. A nonce
is of importance to us because it is very likely to be the common key between many transactions that
cause them to become interdependent. For example, imagine an account \texttt{Alice} that has state
keys in different runtime modules. One for balance, and one for different each module. These state
keys can be accessed independently, but bringing the logic of nonce into the picture, things get a
wee bit more complicated. Now, all of these transactions need to write to the key that stores the
nonce of Alice, and they are all dependent on one another. Things can get even worse. There could be
special state keys that need to be accessed by \textit{all} transactions (for example: a counter for
all transaction in the block that is kept in the state). This forcefully makes all of the
transactions in the block sequential if no special treatment is in place for them. And indeed, this
is our main point: we omit such special cases from our work because they should be treated
\textit{specially} in order for any concurrency model to be sensible.







\bibliographystyle{Latex/Classes/PhDbiblio-url2} 
\renewcommand{\bibname}{References} 

\bibliography{9_references/references} 









\end{document}